\documentclass[twocolumn,trackchanges]{aastex62}

\usepackage{graphicx}
\usepackage{amssymb}
\usepackage{threeparttable}
\usepackage{color}

\shorttitle{Periodic and Phase-locked Modulation in The MP and IP of PSR B1929+10}
\shortauthors{F. F. Kou et al.}

\begin{document}

\title{Periodic and Phase-locked Modulation in PSR B1929+10 Observed with FAST}

\correspondingauthor{F. F. Kou; B. Peng}
\email{koufeifei@nao.cas.cn; pb@nao.cas.cn}

\author[0000-0002-0069-831X]{F. F. Kou}
\affiliation{CAS Key Laboratory of FAST, National Astronomical Observatories, Chinese Academy of Sciences, Beijing 100101, P.\,R.\, China}
\affiliation{Xinjiang Astronomical Observatories, Chinese Academy of Sciences, Urumqi, 830011, P.\,R.\,China}

\author[0000-0002-7662-3875]{W. M. Yan}
\affiliation{Xinjiang Astronomical Observatories, Chinese Academy of Sciences, Urumqi, 830011, P.\,R.\,China}
\affiliation{Key Laboratory of Radio Astronomy, Chinese Academy of Sciences}

\author{B. Peng}
\affiliation{CAS Key Laboratory of FAST, National Astronomical Observatories, Chinese Academy of Sciences, Beijing 100101, P.\,R.\, China}

\author{J. G. Lu}
\affiliation{CAS Key Laboratory of FAST, National Astronomical Observatories, Chinese Academy of Sciences, Beijing 100101, P.\,R.\, China}

\author{K. Liu}
\affiliation{Max-Plank-Institut f{\"u}r Radioastronomie, Auf dem H{\"u}gel 69,
Bonn, D-53121 , Germany}
\affiliation{CAS Key Laboratory of FAST, National Astronomical Observatories, Chinese Academy of Sciences, Beijing 100101, P.\,R.\, China}

\author{C. M. Zhang}
\affiliation{CAS Key Laboratory of FAST, National Astronomical Observatories, Chinese Academy of Sciences, Beijing 100101, P.\,R.\, China}

\author{R. G. Strom}
\affiliation{ASTRON, Postbus 2, 7990 AA Dwingeloo, The Netherlands}
\affiliation{Astronomical Institute, University of Amsterdam, The Netherlands}

\author[0000-0003-0757-3584]{L. Wang}
\affiliation{CAS Key Laboratory of FAST, National Astronomical Observatories, Chinese Academy of Sciences, Beijing 100101, P.\,R.\, China}
\affiliation{Jodrell Bank Centre for Astrophysics, School of Physics and Astronomy\\ The University of Manchester Manchester, M13 9PL, UK}
\affiliation{School of Astronomy and Space Science, University of Chinese Academy of Sciences}

\author{J. P. Yuan}
\affiliation{Xinjiang Astronomical Observatories, Chinese Academy of Sciences, Urumqi, 830011, P.\,R.\,China}
\affiliation{Department of Astronomy, China West Normal University, Nanchong 637009, P.\,R.\,China}

\author{Rai Yuen}
\affiliation{Xinjiang Astronomical Observatories, Chinese Academy of Sciences, Urumqi, 830011, P.\,R.\,China}
\affiliation{Key Laboratory of Radio Astronomy, Chinese Academy of Sciences}

\author{Y. Z. Yu}
\affiliation{Qiannan Normal University for Nationalities, Duyun 558000, P.\,R.\,China}

\author{J. M. Yao}
\affiliation{CAS Key Laboratory of FAST, National Astronomical Observatories, Chinese Academy of Sciences, Beijing 100101, P.\,R.\, China}
\affiliation{Xinjiang Astronomical Observatories, Chinese Academy of Sciences, Urumqi, 830011, P.\,R.\,China}

\author{B. Liu}
\affiliation{CAS Key Laboratory of FAST, National Astronomical Observatories, Chinese Academy of Sciences, Beijing 100101, P.\,R.\, China}

\author{J. Yan}
\affiliation{CAS Key Laboratory of FAST, National Astronomical Observatories, Chinese Academy of Sciences, Beijing 100101, P.\,R.\, China}

\author{P. Jiang}
\affiliation{CAS Key Laboratory of FAST, National Astronomical Observatories, Chinese Academy of Sciences, Beijing 100101, P.\,R.\, China}

\author{C. J. Jin}
\affiliation{CAS Key Laboratory of FAST, National Astronomical Observatories, Chinese Academy of Sciences, Beijing 100101, P.\,R.\, China}

\author{D. Li}
\affiliation{CAS Key Laboratory of FAST, National Astronomical Observatories, Chinese Academy of Sciences, Beijing 100101, P.\,R.\, China}
\affiliation{NAOC-UKZN Computational Astrophysics Centre, University of KwaZulu-Natal, Durban 4000, South Africa}

\author{L. Qian}
\affiliation{CAS Key Laboratory of FAST, National Astronomical Observatories, Chinese Academy of Sciences, Beijing 100101, P.\,R.\, China}

\author{Y. L. Yue}
\affiliation{CAS Key Laboratory of FAST, National Astronomical Observatories, Chinese Academy of Sciences, Beijing 100101, P.\,R.\, China}

\author{Y. Zhu}
\affiliation{CAS Key Laboratory of FAST, National Astronomical Observatories, Chinese Academy of Sciences, Beijing 100101, P.\,R.\, China}


\author{(The FAST collaboration)}

\begin{abstract}
We present a detailed single-pulse analysis for PSR B1929+10 based on observations with the Five-hundred-meter Aperture Spherical radio Telescope (FAST). The main pulse and interpulse are found to be modulated with a periodicity of $\sim12$ times the pulsar's rotational period ($P$). The $\sim12P$ modulation is confirmed as a periodic amplitude modulation instead of systematic drifting. The periodic amplitude modulation in the IP is found to be anti-correlated with that in the weak preceding component of the MP ($\rm MP\_\uppercase\expandafter{\romannumeral1}$), but correlated with that in the first two components of the MP ($\rm MP\_\uppercase\expandafter{\romannumeral2}$), which implies that the modulation patterns in the IP and the MP are phase-locked. What is more interesting is that the modulation in $\rm MP\_\uppercase\expandafter{\romannumeral2}$ is delayed that in the IP by about 1P. Furthermore, high sensitivity observations by FAST reveal that weak emission exists between the MP and the IP. In addition, we confirm that the separation between the IP and the MP is independent of radio frequency. The above results are a conundrum for pulsar theories and cannot be satisfactorily explained by the current pulsar models. Therefore,  our results observed with FAST provide an opportunity to probe the structure of pulsar emission and  the neutron star's magentosphere.
\end{abstract}


\section{Introduction} \label{sec:intro}
To explore pulsar emission geometry and its magnetosphere, 
many observations and studies of PSR B1929+10 (PSR J1932+1059) have been performed because its integrated pulse profile clearly exhibits both a main pulse (MP) and an interpulse (IP) \citep{1990_Phillips_1929,1997_Rankin_1929}, features usually thought to be pulsar emission from two opposite magnetic poles.
PSR B1929+10 was discovered in the Molonglo survey by \citet{1968_Large_pulsar_search}, shown to be a normal pulsar with a characteristic age of $\sim 3.1 \,\rm Myr$, and recently it was also detected in the X-ray band \citep{2006_Becker_1929+10_multiwave,2006_Misanovic_1929_xray}. At radio frequencies, the phase separation between the MP and IP is about $187.4\pm0.2^\circ$ and it is reported to be independent of observing frequency \citep{1985_Perry_unpulsed_emission}.  Polarimetric studies have been carried out at various observing frequencies for this pulsar, showing it has high linear polarization and a very small circular component \citep{1997_Rankin_1929}. According to previous works, the magnetic inclination angle $\alpha$ (the angle between the magnetic axis and the rotation axis), which can be derived by fitting the position-angle of the linear polarization to the rotating vector model (RVM for short), was measured to be approximately $35^\circ$ \citep{1985_Perry_unpulsed_emission,1997_Rankin_1929,1999_Stairs_1929}.

Fluctuation spectrum analysis showed that the pulsar's emission is modulated periodically, with two significant peaks at $0.095\, \rm cycles/period$ (cpp) ($11 P$) and $0.1795\, \rm  cpp$ ($5.57 P$), respectively \citep{1973_Backer_FS,1997_Rankin_1929}. Two drifting features with different periods in opposite directions were found by \citet{2006_Weltevrede} using the Westerbork Synthesis Radio Telescope (WSRT) in the Netherlands at wavelength $21 \,\rm cm$. However, \citet{2016_Basu_Amp_flu,2019_Basu_dirfting} claimed that there is no systematic drifting but periodic amplitude fluctuation with a period of $11.5 \pm1.3 P$ based on Giant Meterwave Radio Telescope (GMRT) observations at $333 \,\rm MHz$.

Weak emission between the MP and the IP was first detected by \citet{1985_Perry_unpulsed_emission}, and was clearly identified at $430\, \rm MHz$ with the Arecibo $300\, \rm m$ telescope \citep{1990_Phillips_1929,1997_Rankin_1929}. Because of the weak emission throughout most of its rotational period,  PSR B1929+10 had been categorized in the group of single-pole interpulse pulsars, which means that the MP and the IP come from the same magnetic pole \citep{1990_Phillips_1929}. However, a contradictory conclusion can be derived based on the pulse profile components. PSR B1929+10 was identified as a two-pole interpulse pulsar according to its core-emission dominated profile \citep{1997_Rankin_1929,2019_Basu_classify}, which means that the MP and the IP come from the two opposite magnetic poles of the pulsar.

In this paper, to  clearly present the emission properties of  PSR B1929+10, we exploited the largest single dish telescope FAST to make single-pulse observations of this pulsar \citep{2011_Nan_FAST}. The high sensitivity observations of PSR B1929+10 with FAST reveal previously unknown and more complex emission properties, which may further complicate our understanding of radio emission in this pulsar. In this paper, we investigate the modulation and interaction between the IP and the MP, and try to discuss the theoretical restrictions based on our results. In section $2$, the observations and data processing method are introduced. The results are given in section $3$. We summarize our results and discuss the possible geometric structure of the magnetosphere for PSR B1929+10 in sections $4$ and $5$, respectively. 

\section{observations}
FAST is located in Guizhou, China, with its whole
aperture being $500\, \rm m$ and with an illuminated sub-aperture of $300\,\rm m$ during normal operation.  A $19$-beam receiver covering $1.05$-$1.45\,\rm GHz$ (the so called  FLAN: \citep{2018_Li_FAST}) has been mounted for use since July $2018$. It contains a temperature stabilized noise injection system, with a noise signal  injected from a single diode \citep{2020_Jiang_19beam}. The collected data are captured by a digital backend based on Reconfigurable Open-architecture Computing Hardware–version$2$ (ROACH$2$)  \citep{2019_JP_FAST,2020_Jiang_19beam}, and are recorded in search-mode PSRFITS data format \citep{2004_Hotan_Psichive}, with time and frequency resolutions of $49.152  \mu s$ and $0.122\,\rm MHz$, respectively.

The observations of PSR B1929+10 in this paper were performed on $2018$ September $5$ and $2019$ November $22$. A total of $1.5$ hours of observations were recorded. More than $20,000$ single pulses were obtained from these two observations. The data were processed to remove dispersion delay caused by the interstellar medium and produce single-pulse archives with the DSPSR software packages \citep{2011_DSPSR}. The ephemeris of the pulsar was obtained from the ATNF pulsar catalogue (PSRCAT, version $1.56$) \citep{2005_Manchester_ATNF}. The channels with strong radio frequency interference (RFI) were removed using the PSRCHIVE packages \citep{2010_PSRCHIVE}. A polarization calibration noise signal was injected at an off-source (pulsar) position and recorded before the pulsar observation on $2019$ November $22$. However, no stable radio flux density calibrator was observed at that time, and no polarization calibration data were recorded for the observation on $2018$ September $5$. We could only get a polarization pulse profile for the observation of $2019$ November $22$.

\section{Results}
\subsection{The single pulse modulation}
\label{LRFS}
\begin{figure}[htp]
\includegraphics[width=0.46\textwidth]{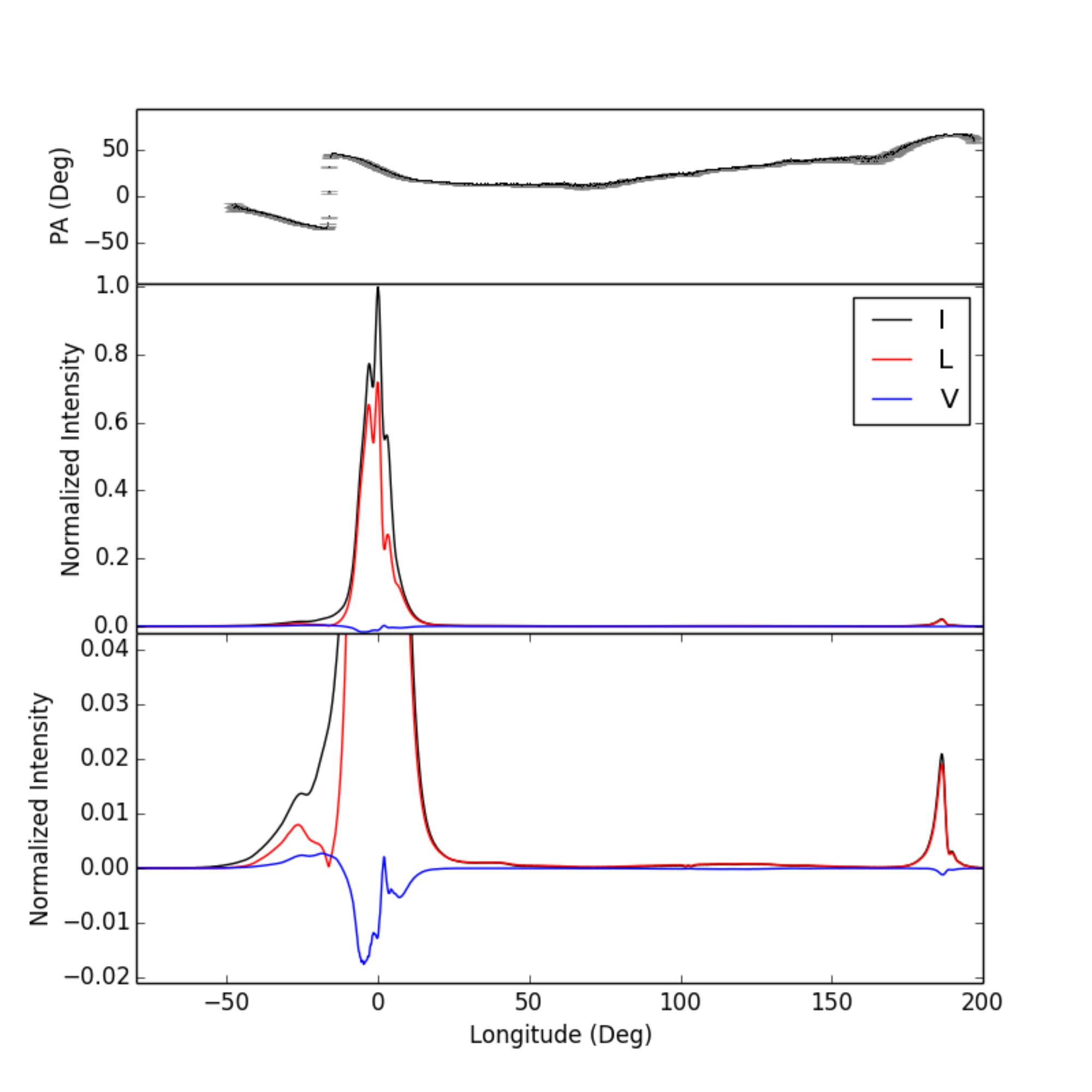}
\caption{The mean pulse profile of PSR B1929+10 from the observation of $2019$ November $22$. The upper panel gives the position angles of the linearly polarized emission. The middle and bottom panels shows the mean pulse profile for total intensity (black line), linearly polarized intensity (red line), and circularly polarized intensity (blue line). The bottom panel is the expanded plot of the middle panel.}
\label{fig:meanpulse}
\end{figure}

\begin{figure}[htp]
\begin{minipage}{0.23\textwidth}
\includegraphics[width=1\textwidth]{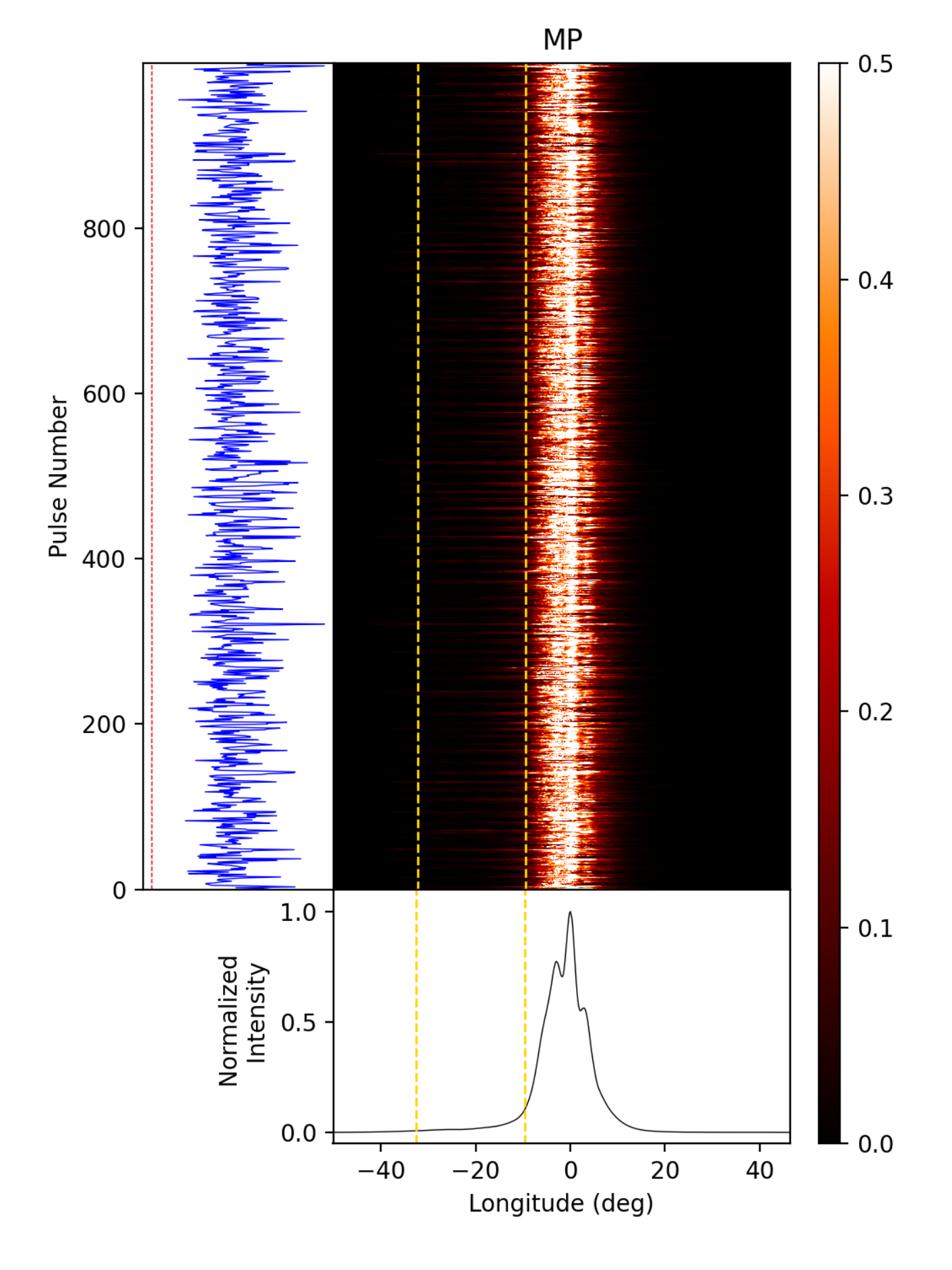}
\end{minipage}
\begin{minipage}{0.23\textwidth}
\includegraphics[width=1.09\textwidth]{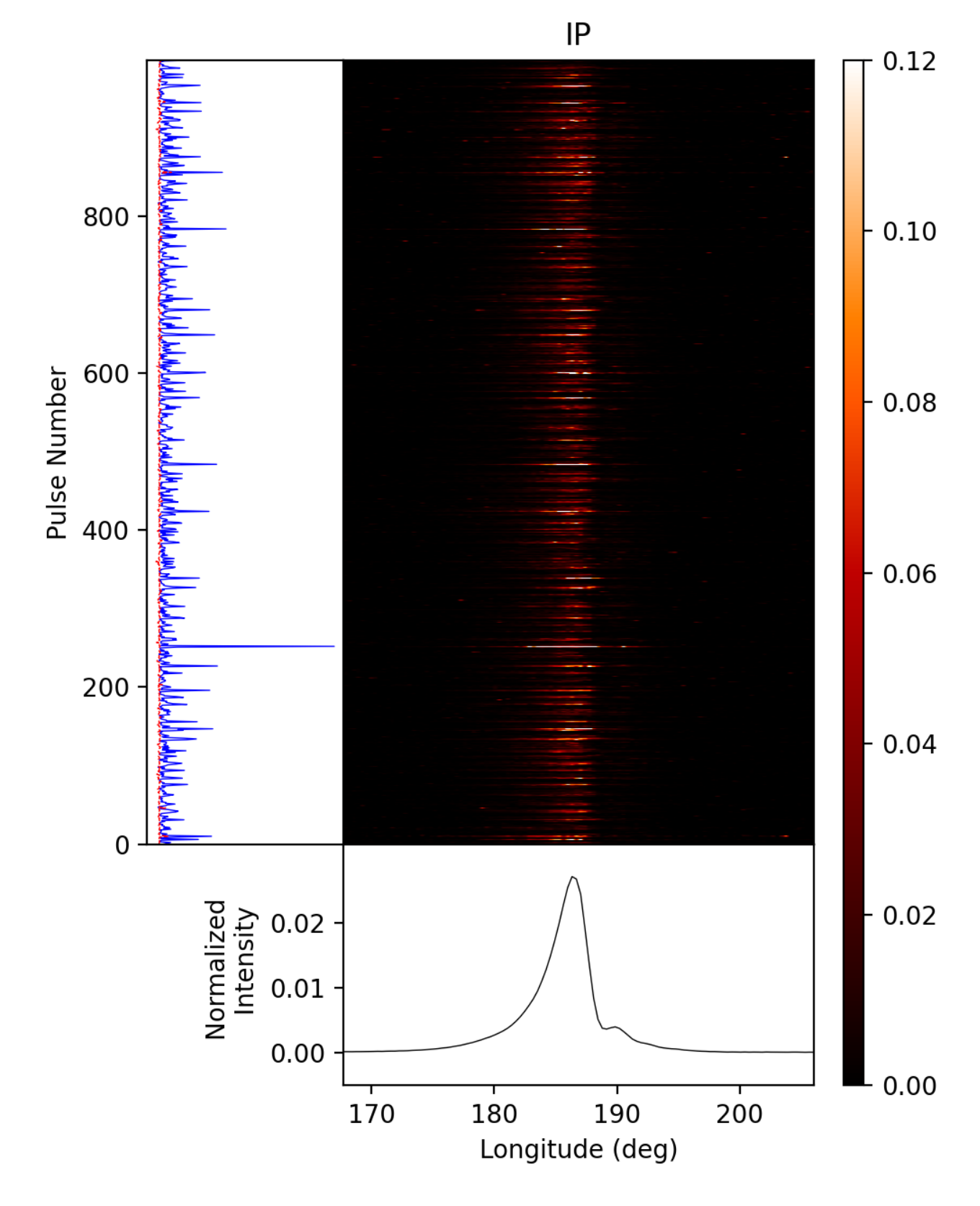}
\end{minipage}
\caption{The single-pulse stack of PSR B1929+10 from the observation of $2019$ November $22$. The left and right columns show the longitude range around the MP and the IP for the same rotation, respectively. The two vertical dashed lines in the left column define the pulse longitudes of the preceding part of the MP. The left panel of each column shows the energy variations for the on-pulse range (blue solid line) and the off-pulse range (red dashed line). The bottom panels show the integrated pulse profiles which are normalized to the peak intensity of the MP .}
\label{fig:sp_stack}
\end{figure}

The polarization profile for the observation on $2019$ November $22$ is shown in Fig. \ref{fig:meanpulse}. The longitude of the MP's pulse peak is set to be zero. The IP follows the MP by $186.3\pm0.2^{\circ}$, with peak intensity about $3\%$ that of the MP at a central frequency of $1.25 \, \rm GHz$. A weak preceding component which is about $20^\circ$ ahead of the MP peak can be identified (see the enlarged plot/bottom panel in Fig. \ref{fig:meanpulse}). A single-pulse stack from $2019$ November $22$ is plotted in Fig. \ref{fig:sp_stack}. The left and right columns show the single pulses of the MP and IP, respectively, for the same rotation of the pulsar. As shown in the right column of Fig. \ref{fig:sp_stack}, it seems that the energy of the IP fluctuates between strong and weak states periodically. The left column shows that the preceding part of the MP has a clearer modulation than other parts of the MP. There is no obvious drifting feature found in the pulse stack.

To investigate the single pulse modulation in detail, we carried out an analysis of fluctuation spectra with the PSRSALSA packages for observations reported here, which contains the modulation index, the Longitude-Resolved Fluctuation Spectrum (LRFS) and  the Two-Dimensional Fluctuation Spectrum (2DFS) \citep{2002_Edwards_2DFS,2006_Weltevrede}. The modulation index is a measure of the factor of  intensity variability from pulse to pulse \citep{2006_Weltevrede}.  The LRFS is used to detect the periodicity of the subpulse modulation. The 2DFS is to determine if the subpulses are drifting in pulse longitude. Here, we chose to plot the vertical axis of the LRFS and 2DFS in units of $P/P_3$, and the horizontal axis of the 2DFS in units of $P /P_2$, where $P_3$ is the vertical drift band separation and $P_2$ is the horizontal drift band separation. 

\begin{figure*}[htp]
\begin{minipage}{0.5\textwidth}
\centering
\includegraphics[width=1.0\textwidth]{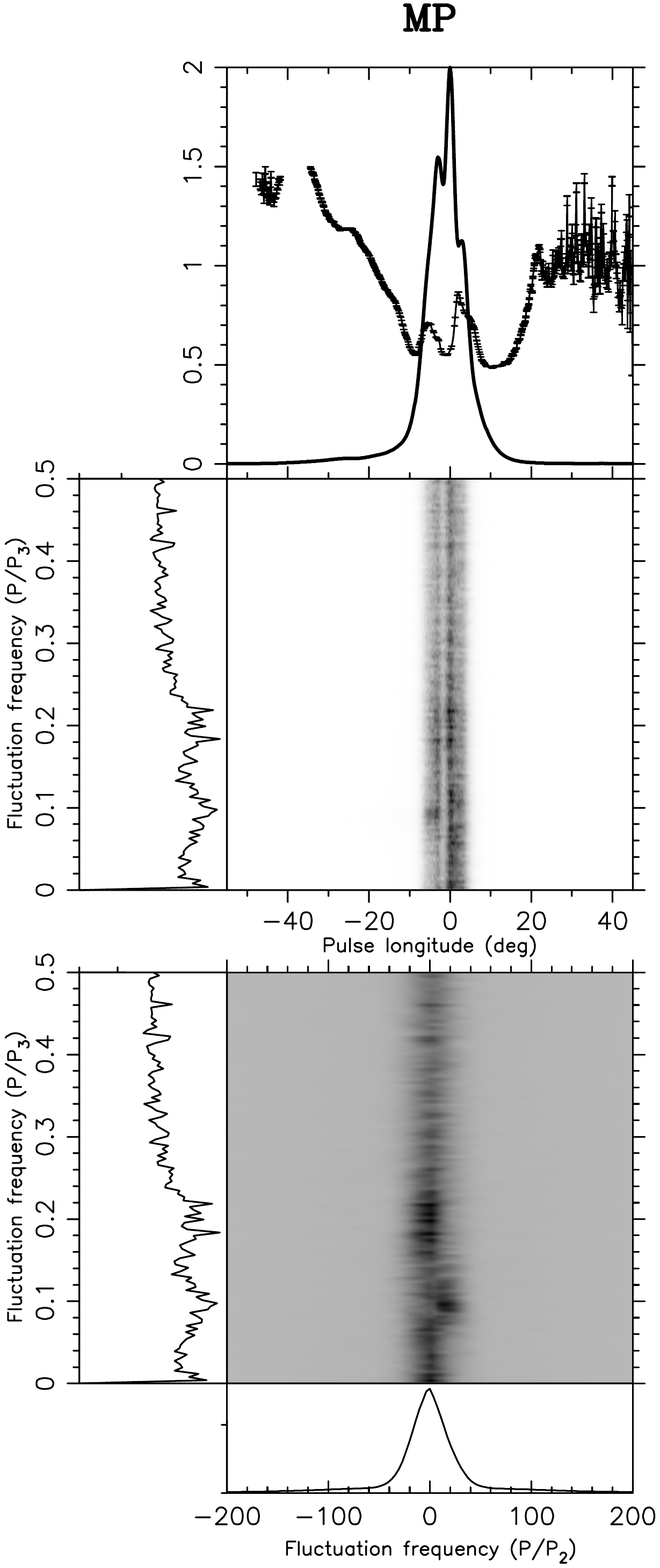}
\end{minipage}
\begin{minipage}{0.5\textwidth}
\centering
\includegraphics[width=1.0\textwidth]{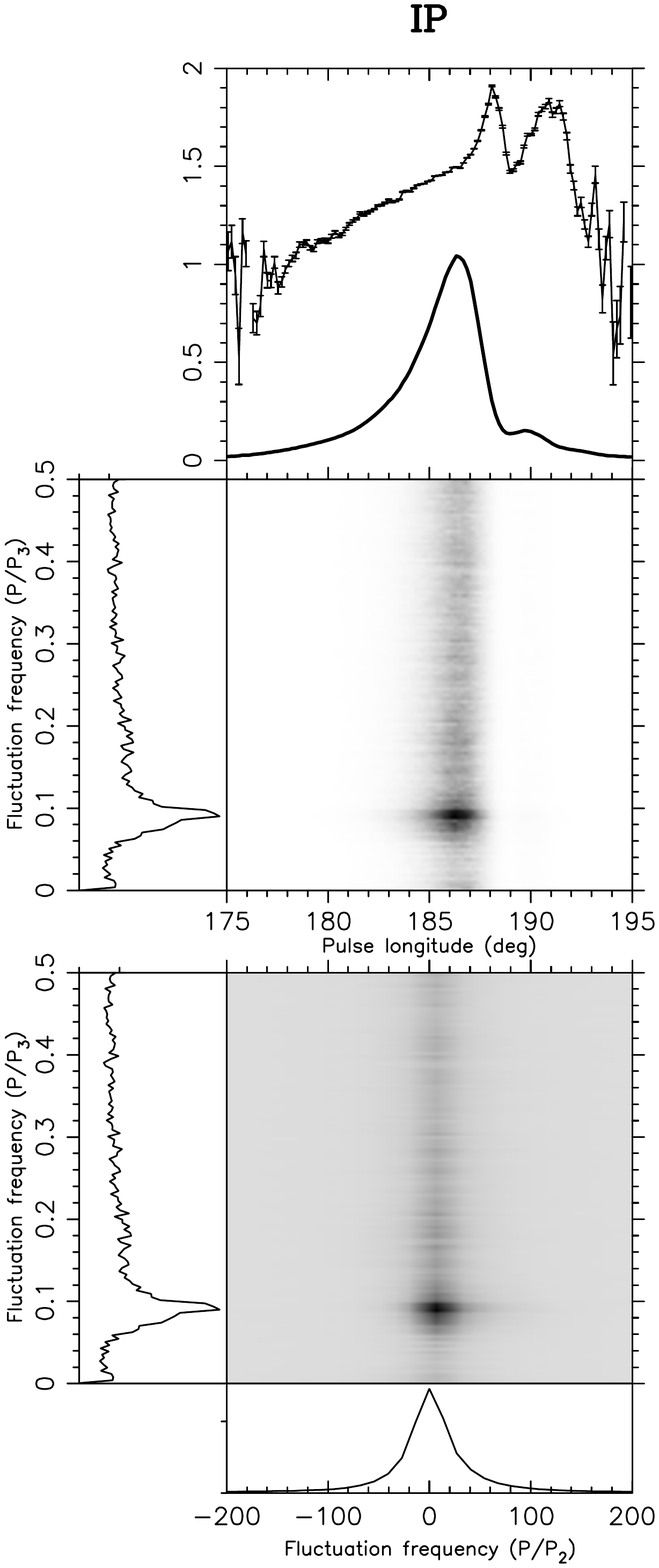}
\end{minipage}
\caption{The fluctuation spectral analysis of PSR B1929+10 for the observation of $2019$ November $22$. The left column is the fluctuation spectrum of the MP for the first $256$ pulses, and the right column is that of the IP. The top panels of each column show the longitude-resolved modulation index and the mean pulse profile.  The intensities of the  MP and the IP are scaled by $2$ times and $50$ times, respectively.  The middle and bottom panels are the LRFS and $2$DFS, respectively.}
\label{fig:FS1}
\end{figure*}

Fig. \ref{fig:FS1} shows fluctuation spectra analyses of PSR B1929+10 based on the observation of $2019$ November $22$. As shown in the top windows of Fig. \ref{fig:FS1}, the modulation indices of the IP and the weak preceding component of the MP are much higher than those of other components. Correspondingly, the spectral analysis of the IP shows clear intensity modulation and identifiable periodicity.
Meanwhile, the spectra of the MP have broad low-frequency features ($0.04\sim 0.22 \, \rm cpp$ ), mainly from the trailing parts. There are two relatively significant features from the preceding part of the MP, with peaks around $0.1 \, \rm cpp$ and $0.18 \, \rm cpp$, which are roughly consistent with the results of \citet{1973_Backer_FS}. \citet{2006_Weltevrede} reported two oppositely drifting features with different $P_3$ values, and they indicated that those structures arose from the leading half of the MP. However, the 2DFS (bottom windows of Fig. \ref{fig:FS1}) shows no obvious horizontal feature, which indicates that there is no systematic drifting but only periodic amplitude fluctuation in our data.

\begin{figure*}[htp]
\begin{minipage}{0.5\textwidth}
\centering
\includegraphics[width=0.7\textwidth]{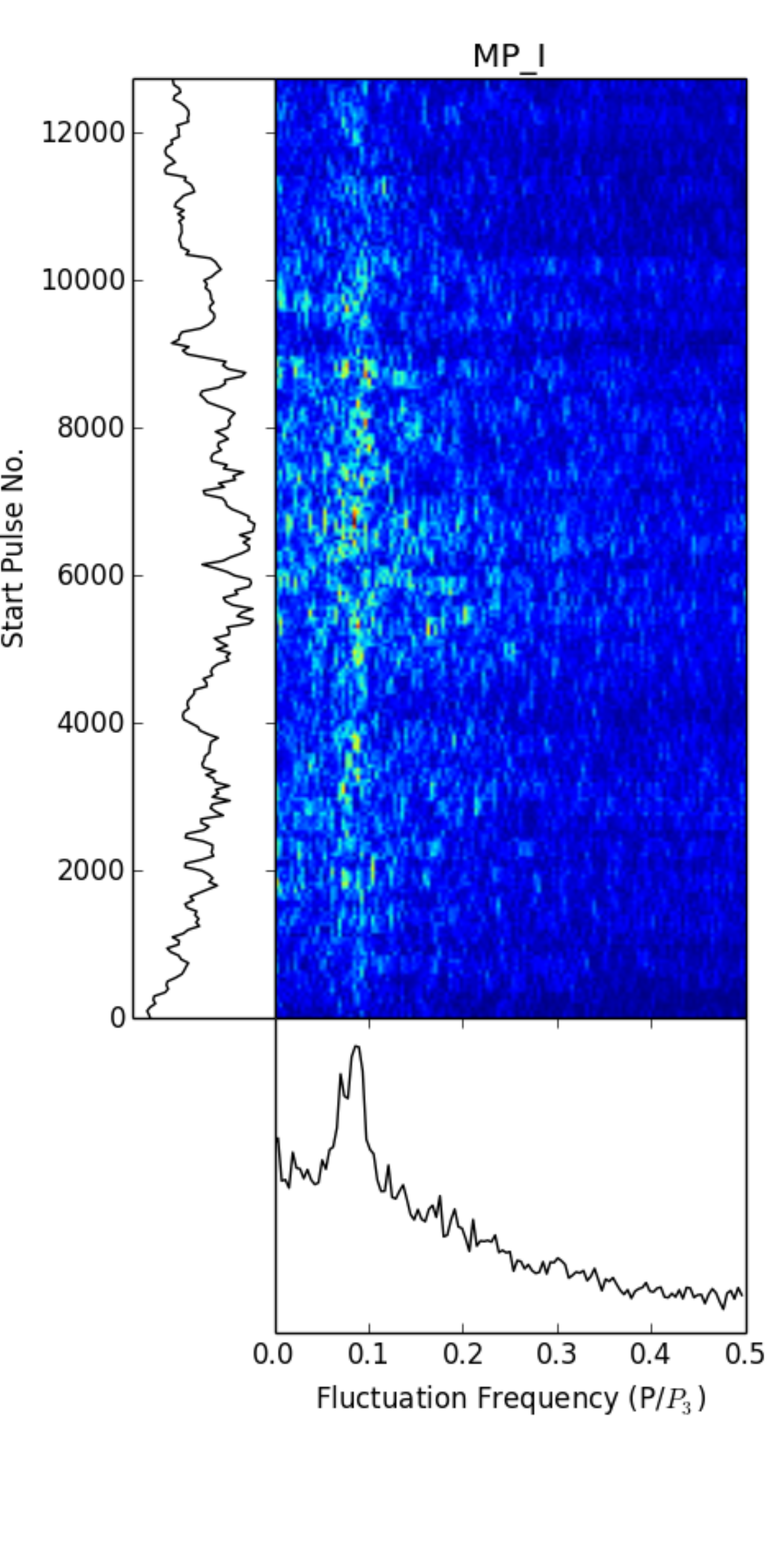}
\end{minipage}
\begin{minipage}{0.5\textwidth}
\centering
\includegraphics[width=0.7\textwidth]{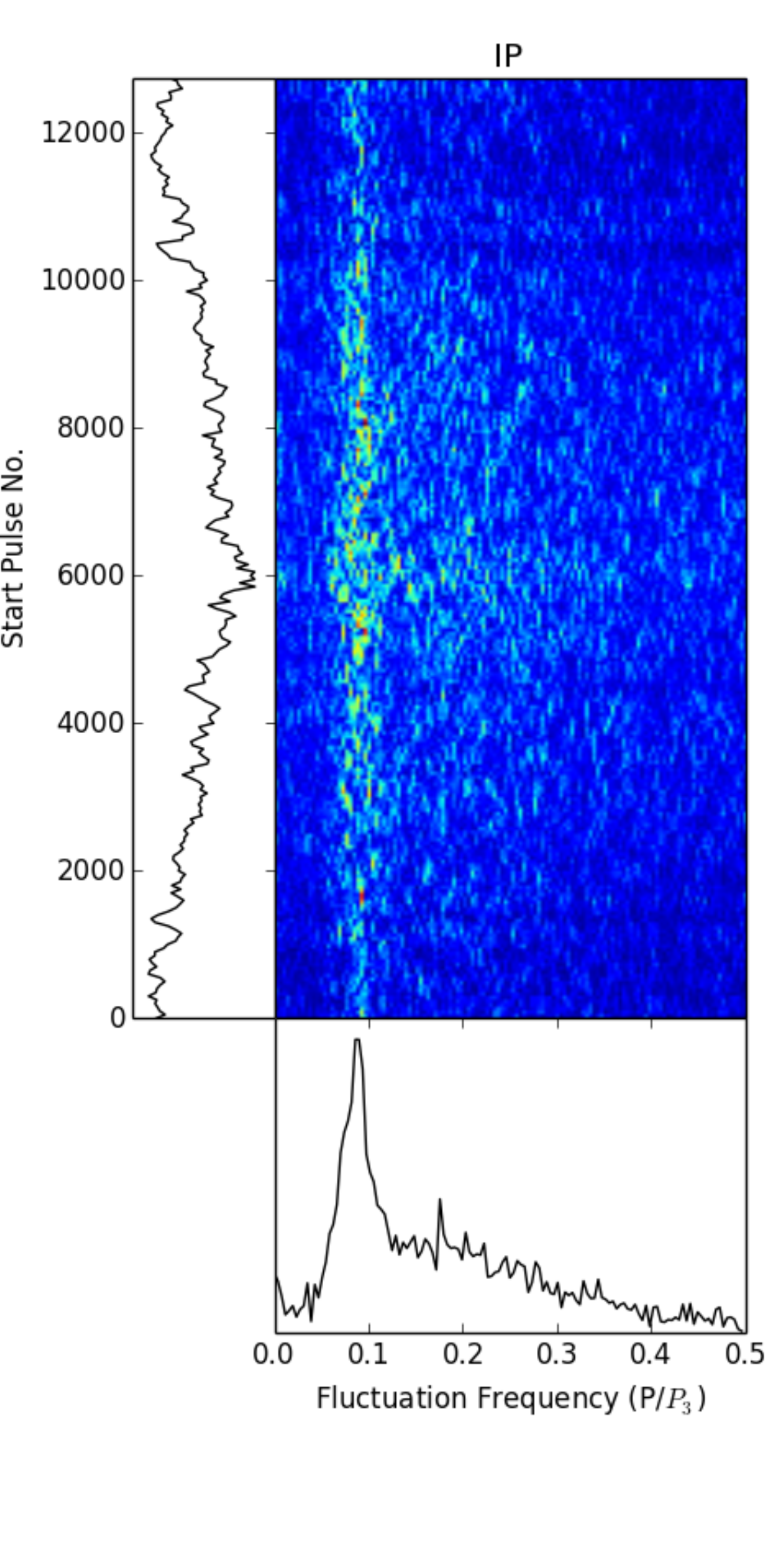}
\end{minipage}
\caption{The time varying LRFSs of the weak preceding component of the MP (left column, labeled as $\rm MP\_I$) and the IP (right column). The side panel of each column shows the temporal variation of the LRFS, and the bottom panel is the average LRFS.}
\label{fig:FS2}
\end{figure*}

The time varying LRFS was used to check the stability of the fluctuation \citep{2016_Basu_Amp_flu}. This was done by calculating the LRFS for each $256$-pulse block of the entire observation by shifting the starting point by $50$ periods.  
In order to avoid being affected by the relatively strong MP emission, the spectrum of its weak preceding component ($\rm MP\_\uppercase\expandafter{\romannumeral1}$), from $-45^{\circ}$ to $-16^{\circ}$, was analysed separately. The time variation of the LRFS of the IP and the weak preceding component is shown in the bottom row of Fig. \ref{fig:FS2}. From the average LRFS, the peak frequency $f_{p}$ and the corresponding error $\delta{f_{p}}$ could be determined \citep{2016_Basu_Amp_flu}. It is $0.086\pm0.014 \, \rm cpp$ for the IP and $0.086\pm0.018  \, \rm cpp$ for the weak preceding component, which implies a modulation period of $P_{3}\sim 12 P$ for both the IP and the weak preceding components.  
Our results are consistent with those of \citet{2016_Basu_Amp_flu} who used GMRT data at $333 \, \rm MHz$. This implies that the fluctuation features could be independent of frequency. 

\subsection{The Phase Locking}
\label{phaselock}
\begin{figure}[htp]
\includegraphics[width=0.46\textwidth]{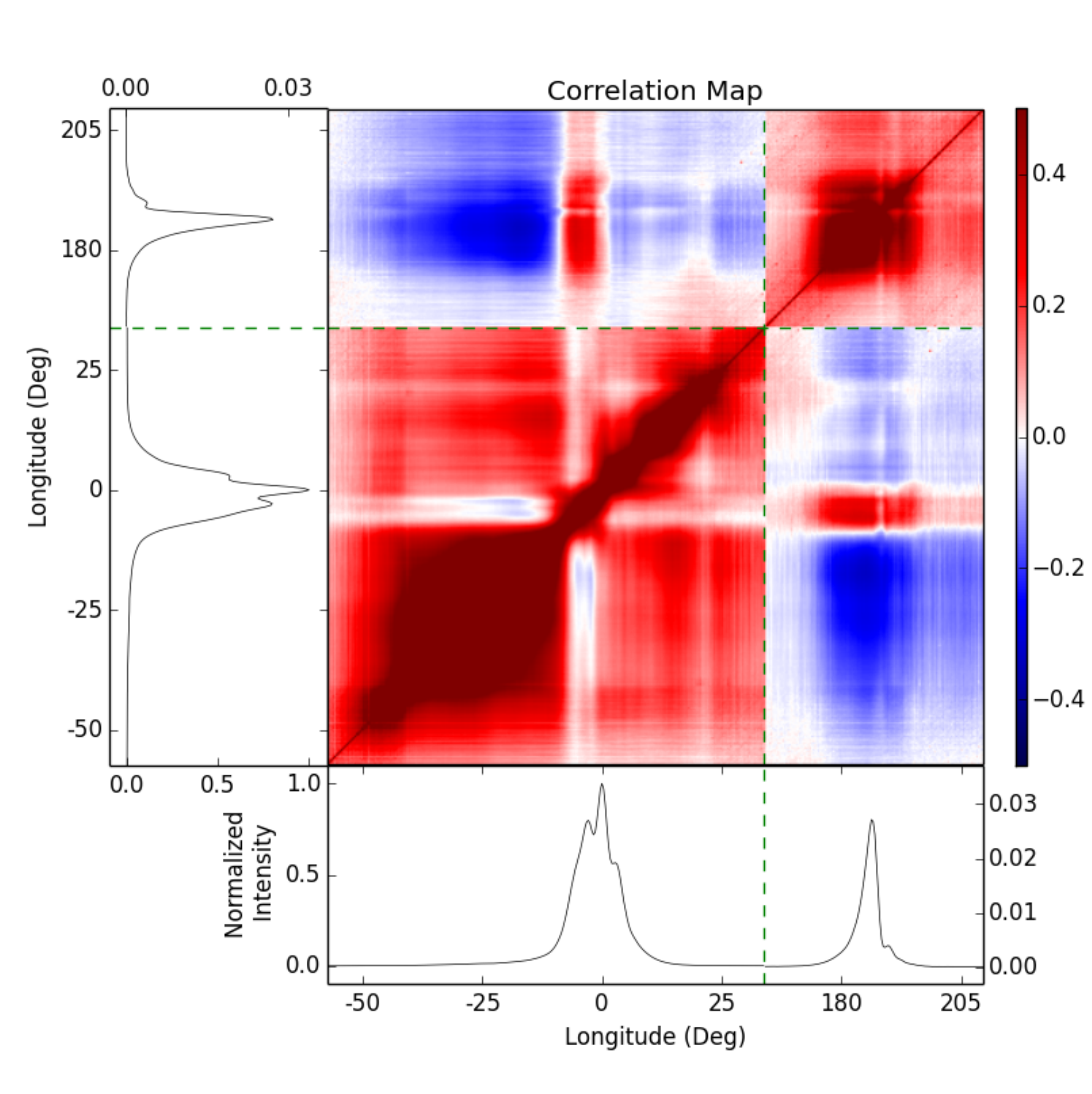}
\caption{The longitude-longitude correlation between the MP and the IP for zero delay. The side and bottom windows show the mean pulse profile with intensity normalized to the peak of the MP. Longitudes between the MP and the IP (from $34^\circ$ to $164^\circ$) are deleted. The green dashed lines defined as the end longitude of the MP ($34^\circ$) and the beginning longitude of the IP ($164^\circ$). }
\label{fig:correlation}
\end{figure}

Since the IP and the weak preceding component of the MP show similar periodic fluctuations, we explore the phase connection between them in this subsection.  The longitude-longitude correlation between the MP (from $-57^\circ$ to $34^\circ$) and the IP (from $164^\circ$ to $210^\circ$)  at zero delay is shown in Fig. \ref{fig:correlation}. 
The square regions on the diagonal are the autocorrelations for the MP and the IP. 
The rectangles on the top-left and bottom-right are the cross correlation regions between the MP and the IP. The region is largely negative for the correlation (blue) between the IP and the weak preceding component of the MP, but weakly positive (red) for the correlation between the IP and the first two components of the MP. This means that the emission in IP is anti-correlated with that of the weak preceding component of the MP, but correlated with that of the first two components of the MP. As discussed in subsection \ref{LRFS}, there are also two relatively significant modulation features in the first two components of the MP with peaks around $0.1 \, \rm cpp$ and $0.18 \, \rm cpp$  (the left column of Fig. \ref{fig:FS1}). The correlation diagram indicates that the pulse longitudes with the same modulation period are locked.

\begin{figure}[htp]
\includegraphics[width=0.46\textwidth]{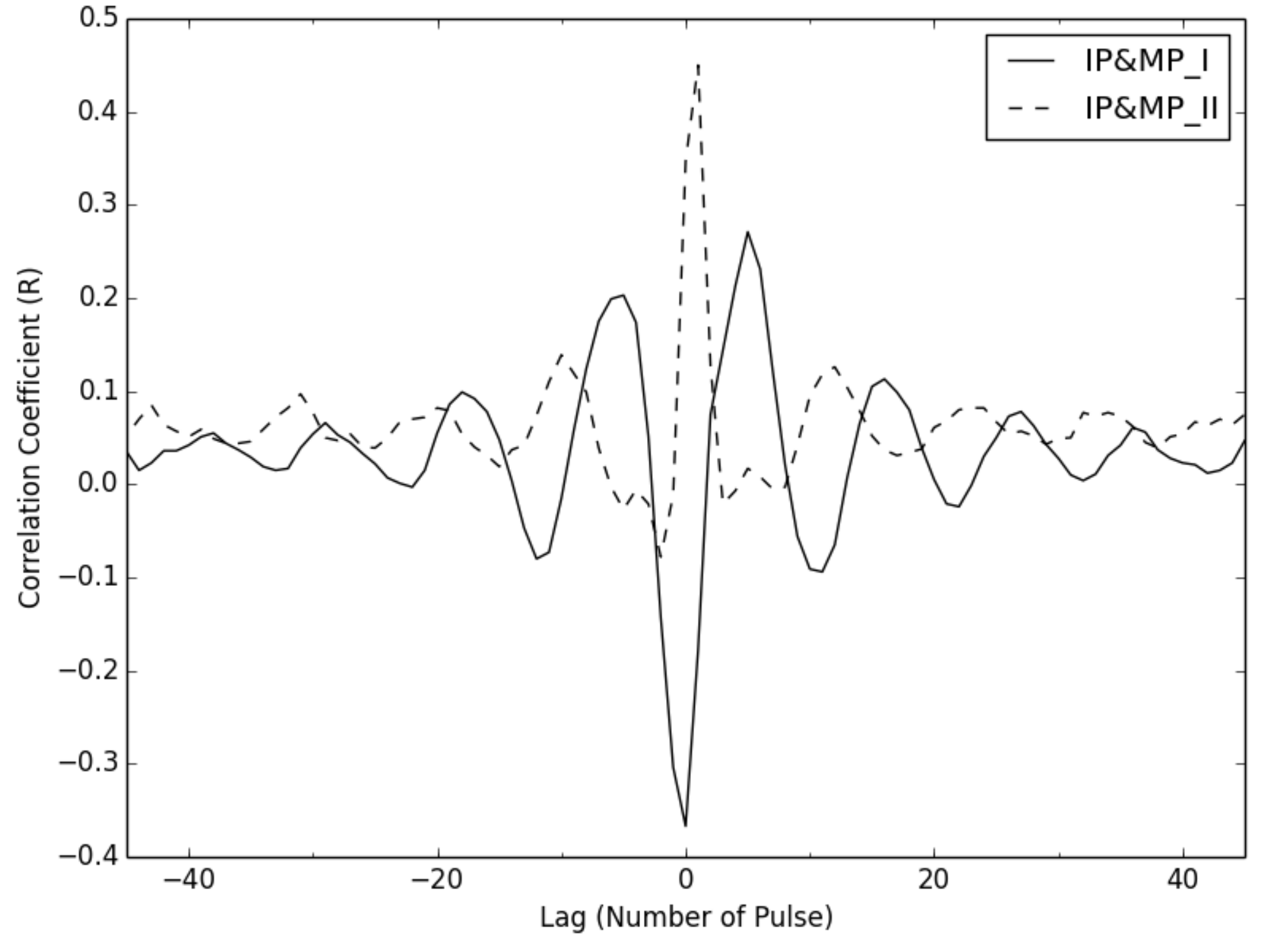}
\caption{The cross-correlations between the pulse energies of the IP and other parts of the MP.  The solid line is for the IP and the $\rm MP\_I$ , and dashed line is for the IP and the $\rm MP\_II$.}
\label{fig:correlation_R}
\end{figure}

According to the correlation diagram, the longitudes of the MP with negative correlation correspond to the weak preceding component ($\rm MP\_\uppercase\expandafter{\romannumeral1}$). The regions with positive correlation correspond to the first two components (from $-15^{\circ}$ to $2^{\circ}$), and called $\rm MP\_\uppercase\expandafter{\romannumeral2}$. The cross-correlations between the pulse energies of the IP and parts of the MP are shown in Fig. \ref{fig:correlation_R}. It can be seen that the cross-correlation reaches its maximum value of $-0.37$ at zero lag (the solid line). This means that the pulse energy modulation of the IP is anti-correlated with that of the $\rm MP\_\uppercase\expandafter{\romannumeral1}$, which is called phase-locked modulation. Furthermore, the $\sim 12 P$ modulation feature in the IP and the $\rm MP\_\uppercase\expandafter{\romannumeral1}$ is also clearly detected from the cross-correlation function. Contrary to the negative correlation between the IP and the $\rm MP\_ \uppercase \expandafter{\romannumeral1}$, the energy variation of the IP is positively correlated with that of the $\rm MP\_\uppercase \expandafter{\romannumeral2}$ (the dashed line in Fig. \ref{fig:correlation_R}). The peak of the correlation function is about $0.45$. We notice that the peak of the correlation function is offset from zero lag, which corresponds to a delay of $\sim 1 P$. This is referred to as a phase-locked delay \citep{2012_Weltevrede_1055}. 

\begin{figure*}[htp]
\begin{minipage}{0.6\textwidth}
\centering
\includegraphics[width=0.99\textwidth]{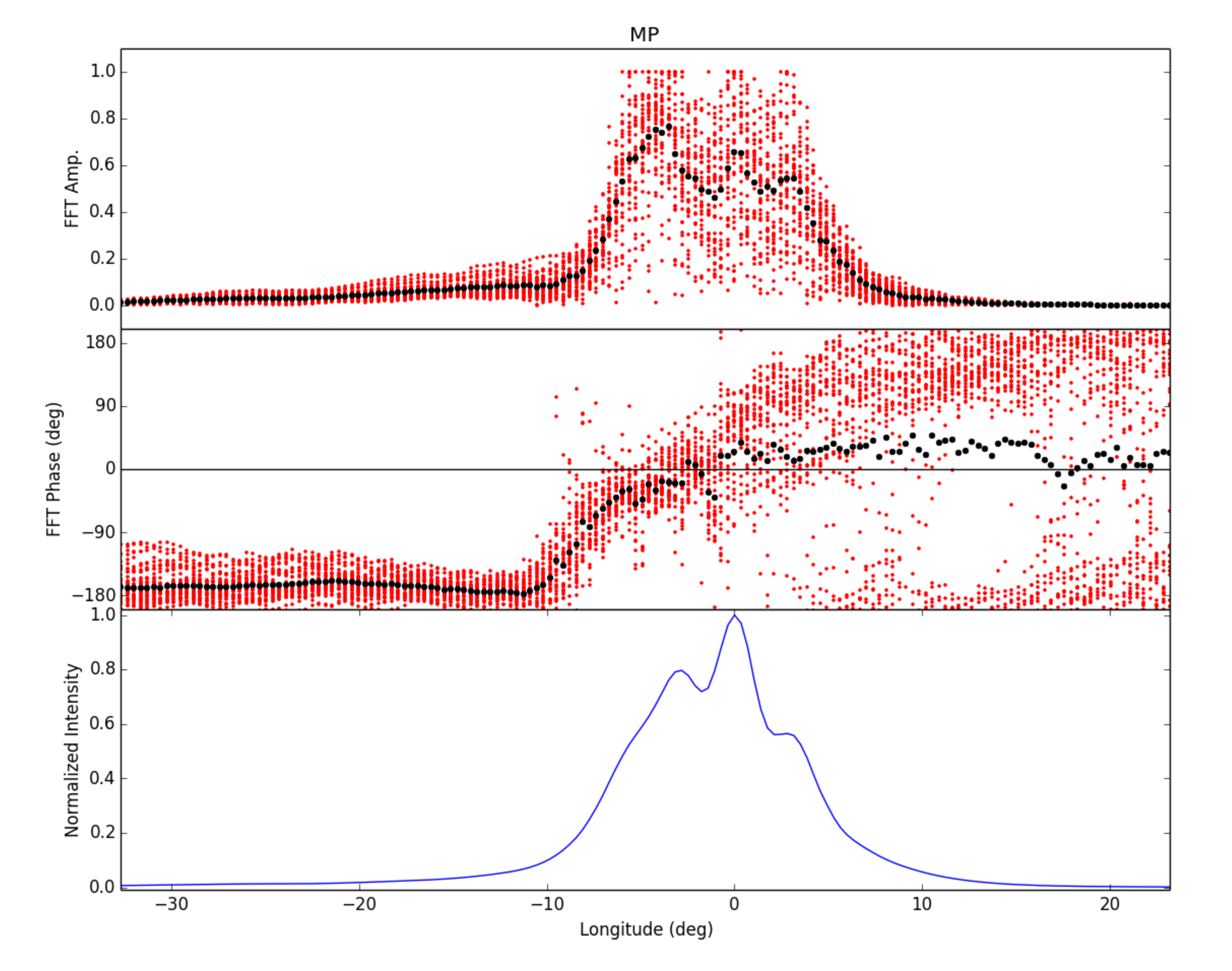}
\end{minipage}
\begin{minipage}{0.4\textwidth}
\centering
\includegraphics[width=0.875\textwidth]{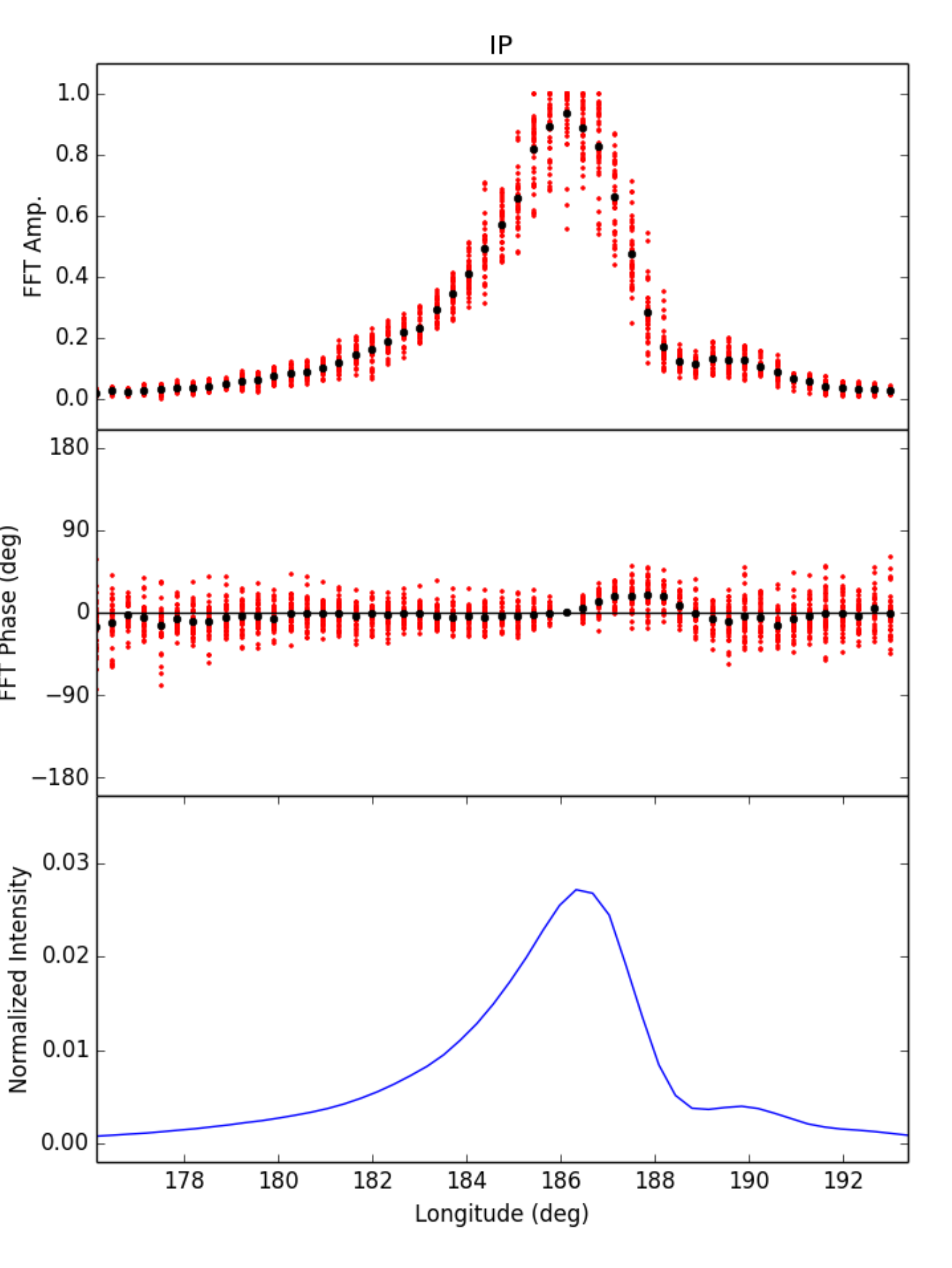}
\end{minipage}
\caption{Variations of the LRFSs across the pulse window for the MP (left column) and the IP (right column) of PSR B1929+10. The panels from top to bottom show the peak amplitudes, the corresponding phase variations and the normalized integrated pulse profiles. The peak amplitudes and the phases for each LRFS are shown as red points, while their average values are shown as black points.}
\label{fig:Phase}
\end{figure*}

We carried out further analysis of the phase variation to study the $\sim12 P$ modulation in detail, which is shown in Fig. \ref{fig:Phase}. Based on the time varying LRFS, the fluctuation spectra with strongest $\sim12 P$ modulation were selected\footnote{The fluctuation spectra value at the frequency corresponding to the  $\sim12 P$ modulation is $3$ times larger than the rms level of the baseline.}. The spectral amplitudes at the frequency of  the  $\sim12 P$ modulation were calculated and averaged in each pulse bin. The amplitudes and their averaged values are shown as red points and black points in the top panels of Fig. \ref{fig:Phase}, respectively. The corresponding phase variations were also estimated. To avoid arbitrary phase differences between different blocks of LRFSs, the phase at the pulse longitude of the IP's peak intensity was set to be zero, and the phase differences across the pulse window were estimated. The phase differences and their averages at a given phase longitude are plotted as red points and black points in the middle panels of Fig. \ref{fig:Phase}, respectively. As shown in the right column of Fig. \ref{fig:Phase}, the amplitudes during the on-pulse range of the IP change consistently, and the corresponding phase differences are close to $0^\circ$. This confirms that the whole IP is modulated by the same period ($\sim 12 P$). Moreover, there is no significant drifting feature detected from the phase differences. For the MP, as shown in the top-left panel of Fig. \ref{fig:Phase}, the distributions of the amplitudes of  $\rm MP\_\uppercase\expandafter{\romannumeral1}$ are more concentrated than those of other parts of the MP, indicating that the whole weak preceding component has the same modulation period. It also shows that the $\sim12 P$ modulation of the MP comes mainly from the weak preceding component. The phase differences of $\rm MP\_\uppercase\expandafter{\romannumeral1}$ are concentrated at $-180^{\circ}$, meaning that the phase of $\rm MP\_\uppercase\expandafter{\romannumeral1}$ is offset with respect to that of the IP by $-180^{\circ}$. The corresponding time delay is about $0.5\, P_3$ (or $\sim6\,P$) between the modulation pattern of the $\rm MP\_\uppercase\expandafter{\romannumeral1}$ and the IP, which confirms the anti-correlation between pulse energies of the $\rm MP\_\uppercase\expandafter{\romannumeral1}$ and the IP.

The phase differences of the first two components of the MP ($\rm MP\_\uppercase\expandafter{\romannumeral2}$) decrease. The difference is $\sim-30^\circ$ for the central region of $\rm MP\_\uppercase\expandafter{\romannumeral2}$, corresponding to a time delay of $\sim1\,P$. 
 This results in the $\sim1\,P$ delay in the modulation of  $\rm MP\_\uppercase\expandafter{\romannumeral2}$ (the dashed line in Fig. \ref{fig:correlation_R}). It is noticed that there is phase evolution between the $\rm MP\_\uppercase\expandafter{\romannumeral1}$ and $\rm MP\_\uppercase\expandafter{\romannumeral2}$. Consequently, the cross-correlation between them and the IP changes from negative to positive. Moreover, this correlation change is accomplished in a few pulse bins (less than $2^\circ$ in pulse phase), so there is almost no transition process. However, for other parts of the MP, the $\sim 12\, P$ modulation pattern is weak, and the phase is dispersedly distributed. We conclude that the phases with strong $\sim 12 P$ modulation features in the IP and the MP are phase locked. The same phase-locked (delay) modulations were also detected in the observations of 2018 September $5$.

\subsection{The emission states and ``bridge'' emission}
\begin{figure}[htp]
\includegraphics[width=0.45\textwidth]{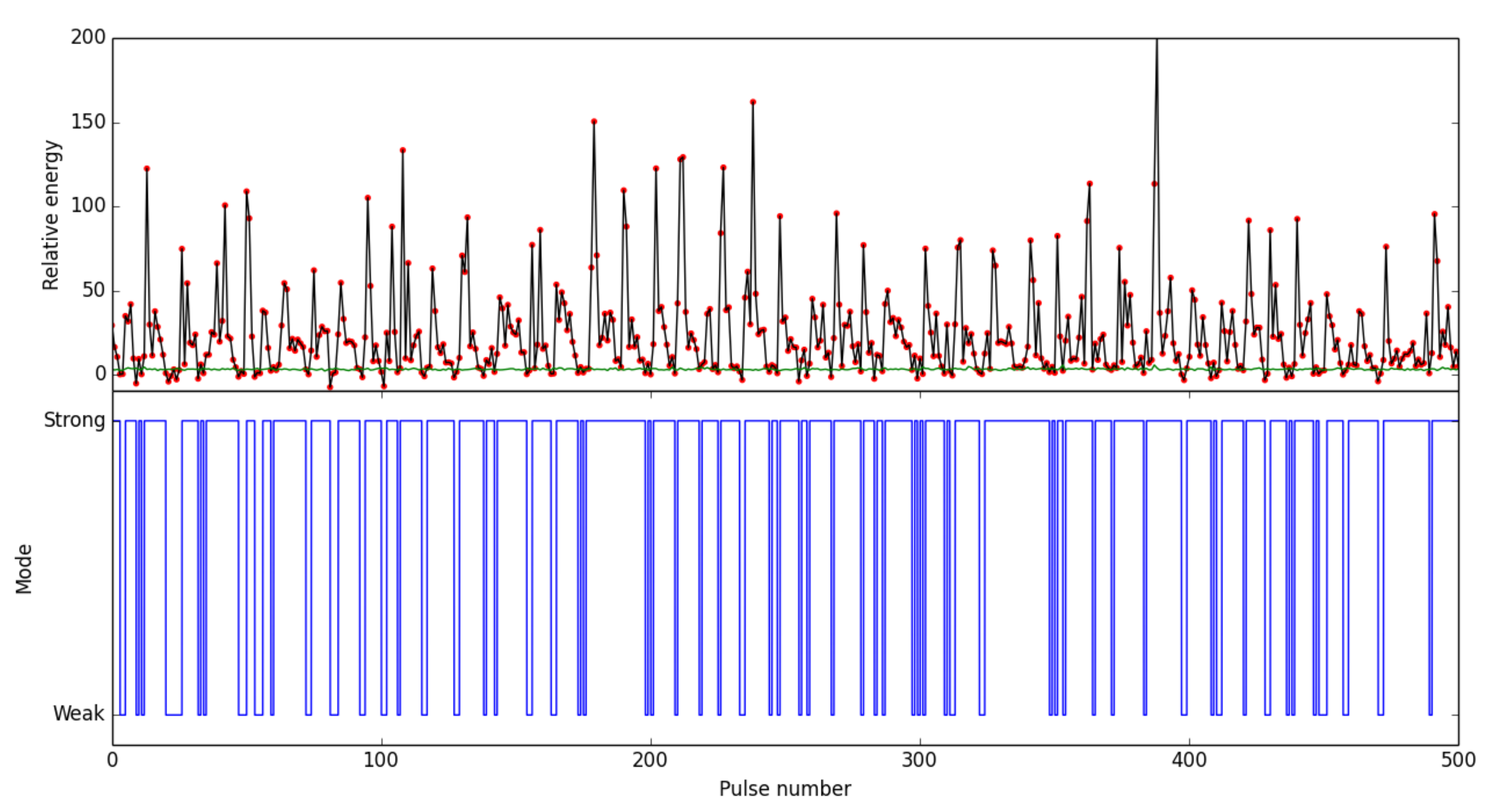}
\caption{The pulse energy sequence of the IP (upper panel) and  the separated  emission modes (bottom panel). The red points in the upper panel are the on-pulse energies of the IP and the green line is the $3\sigma_{\rm off}$ level of the off-pulse region for single pulses. Pulses with energy below the green line are classified as being in the weak mode, and others are taken as the strong mode.}
\label{fig:modes}
\end{figure}

\begin{figure*}
\centering
\includegraphics[width=1.0\textwidth]{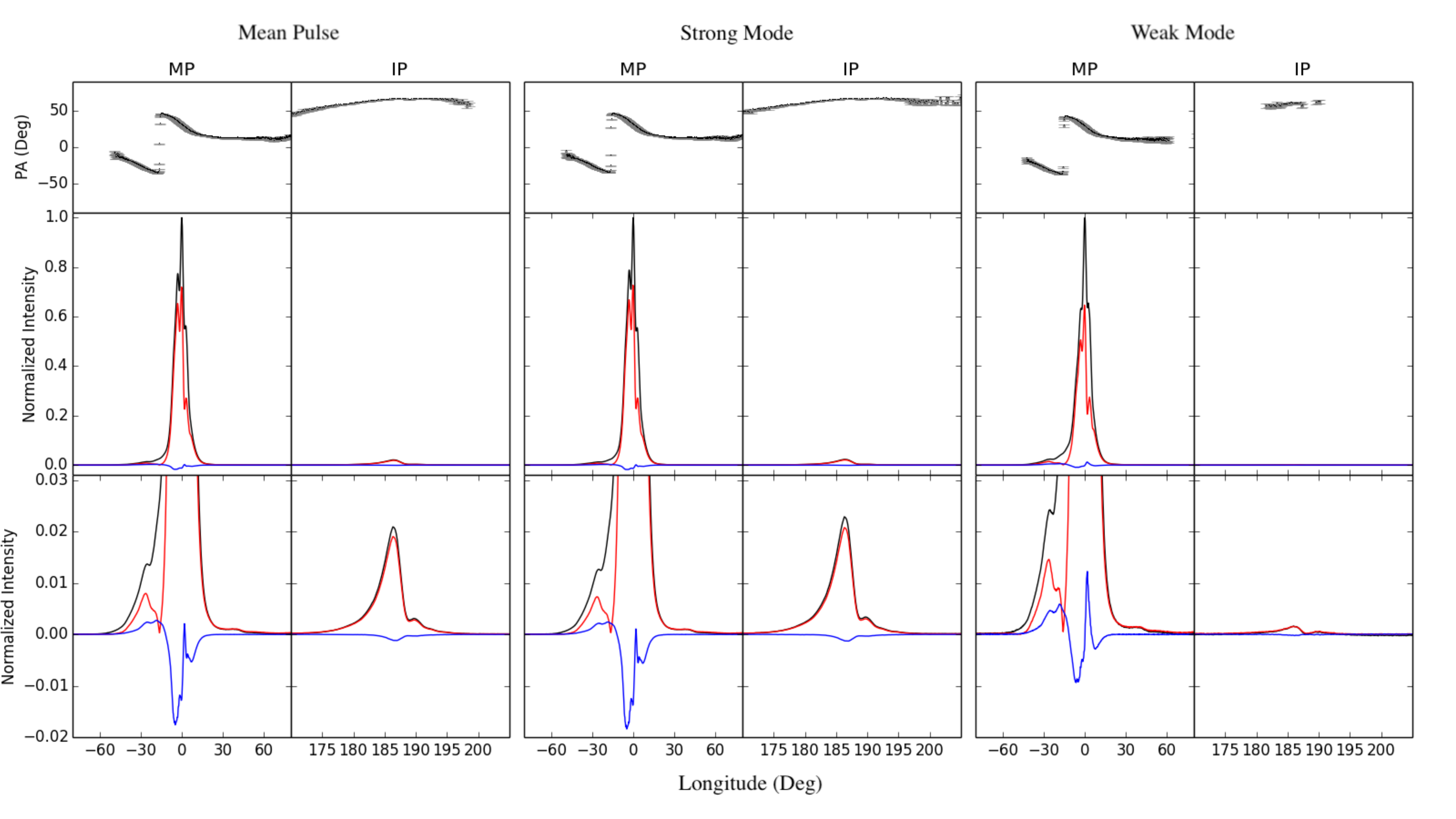}
\caption{The polarized pulse profiles of PSR B1929+10 for the mean pulse (left), the strong mode (middle) and the weak mode (right) from the observation of 2019 November $22$. The top panels show the position angles of the linearly polarized emission. The middle panels show the total intensity (black), the linearly polarized intensity (red) and the circularly polarized intensity (blue). The bottom panels are the expanded plots of the middle panels.}
\label{fig:modes_2}
\end{figure*}

In this subsection, the emission modes are identified according to the periodic energy variation of the IP. Single pulses with on-pulse energy of the IP larger than $3\sigma_{\rm IP, on}$ are classified as strong-mode pulses. The others are weak-mode pulses, where $\sigma_{\rm IP, on}$ is the uncertainty of the on-pulse energy of the IP. $\sigma_{\rm IP, on}=\sqrt{N_{\rm on}}\sigma_{\rm off}$ is the uncertainty in the on pulse energy of the IP, where $N_{\rm on}$ is the number of on-pulse bins and $\sigma_{\rm off}$ is the rms of the off-pulse range \citep{2010_Bhattacharyya,2019_Yan_1822}. The pulse energy sequence of the IP and the separated emission modes are given in Fig. \ref{fig:modes}. The number ratio of the strong mode pulse to the weak mode one is about $6$ to $1$. The polarization profiles for different states are shown in Fig. \ref{fig:modes_2}. The strong mode has a strong IP and weak preceding component of the MP, while the weak mode is the opposite. It is noticed that the interpulse-like structure is also visible in the weak mode, which implies that the periodic longitude intensity modulation is not periodic pulse nulling. A summary of the polarization parameters is given in table \ref{parameter}, where the mean total intensity $\rm I$, the mean linear polarization intensity $\rm\langle L \rangle$ and the mean circular polarization intensity $\rm\langle \vert V \vert \rangle$ were all averaged over the on-pulse windows of the MP and the IP. 
As detected in previous studies \citep{1997_Rankin_1929}, PSR B1929+10 has high linear polarization, especially for the IP, whose linearly polarized intensity exceeds $90\%$ of its total intensity. The fraction of the linear polarization and the circular polarization of the IP in the weak mode are higher than that in the strong mode. Due to the periodic modulation of $\rm MP\_\uppercase\expandafter{\romannumeral2}$, the pulse width of the MP also show periodic changes. As reported in previous works, the polarization position angle shifts near the leading edge of the main pulse. This corresponds to a null in the linear polarized intensity, which can be explained by a switch in two competing orthogonal emission modes \citep{1984_Stinebring_OPM}.

By fitting the RVM to the position angle,  
 a small inclination angle of $\sim 35^\circ$ was found for PSR B1929+10 \citep{1997_Rankin_1929}.  
However, a relatively large inclination angle of $61^\circ$ was given by \citet{1999_Stairs_1929}. In this work, we have carried out an RVM fit to the PA swing using the program of PSRSALSA.  The best RVM fit to the PA swing for the whole phase (with longitudes from $-50^\circ$ to $200^\circ$) shows that the inclination angle $\alpha$ is $145.3^\circ$ and the impact angle of the sight line is $-20.2^\circ$ for the MP and $91^\circ$ for the IP. The best fit has a reduced $\xi^2=16628$. We then only select the longitudes of the MP for the RVM fitting (with longitudes from $-50^\circ$ to $50^\circ$), the best fit $\alpha$ and $\beta$ are $133.5^\circ$ and $-24.0^\circ$ with reduced $\xi^2=29.9$.  It is still not a good fit because the position angle swing contains many structures and deviates from the typical ``S'' shape. We will do further analysis on the single-pulse polarized features for such sources in the following work.

\begin{figure}
\includegraphics[width=0.5\textwidth]{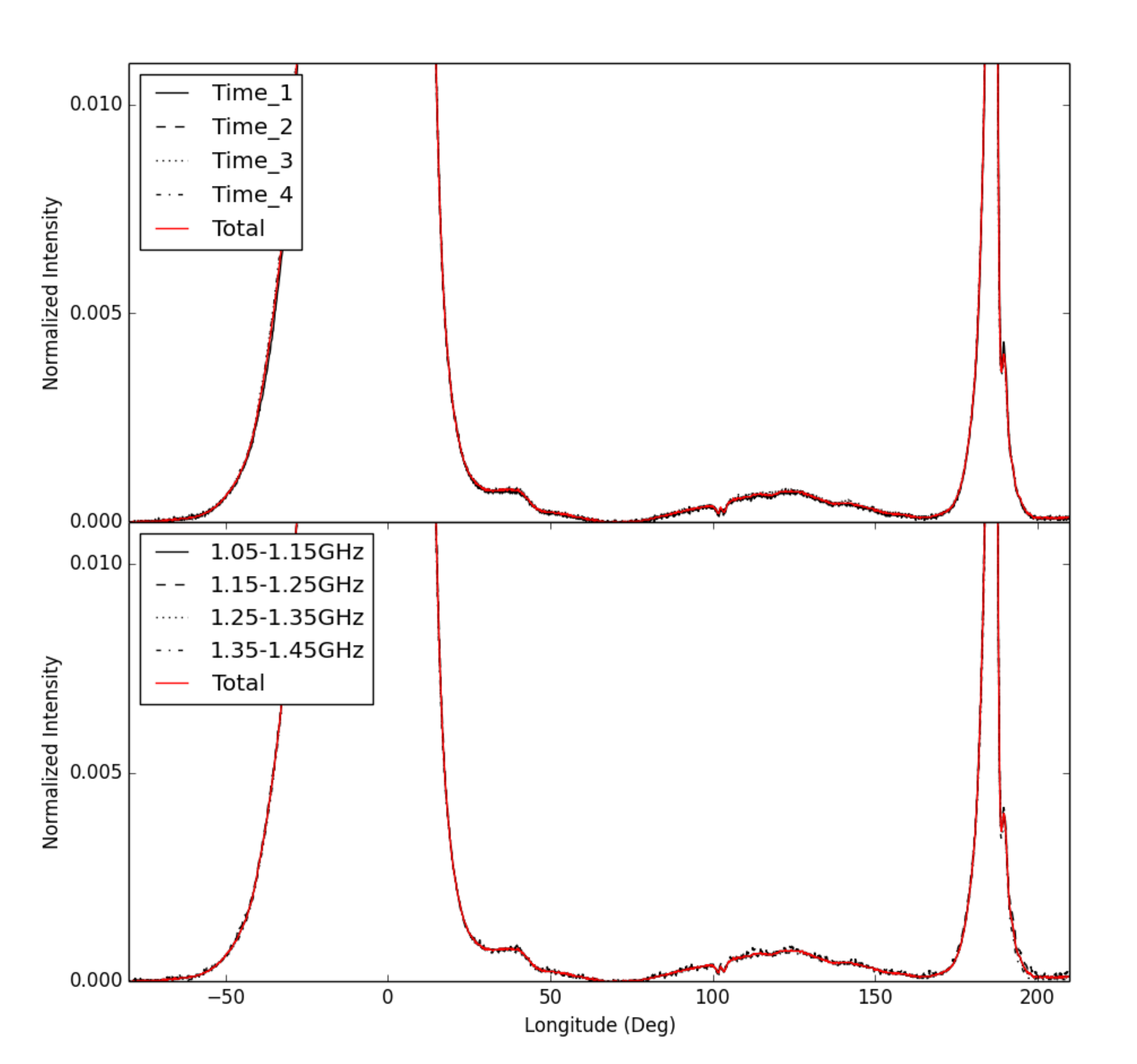}
\caption{The integrated pulse profiles at different time and frequency intervals. The intensity is normalized to the peak intensity of the MP.}
\label{fig:bridge}
\end{figure}

Weak emission following the MP (or preceding the IP) was clearly detected at $430\, \rm MHz$ by the Arecibo telescope, which was reported to have high fractional linear polarization just like the MP and the IP \citep{1997_Rankin_1929}. However, it was comparable to the noise level at $1414\, \rm MHz$ in previous studies. The weak emission following the MP is obviously detected in our data. In order to avoid being affected by random noise, the observation was divided into several intervals in both time and frequency. Fig. \ref{fig:bridge} shows the integrated pulse profiles at different time and frequency intervals. In general, the low-level emission in pulse longitudes from $ 70^\circ$ to $165^\circ$ is visible in all of the individual data sets, like a ``bridge'' connecting the MP and the IP and called postcursor component (PC for short) in previous works \citep{1988_Lyne_beam,1997_Rankin_1929}. The fractional linear polarization of this component is almost $\sim 100\%$.  We carried out fluctuation spectral analysis for the ``bridge'' emission, but there were no obvious modulation features found. However, there is no such ``bridge'' emission on the other side of the IP. As shown in the Fig. \ref{fig:bridge}, its position is also independent of frequency.

\begin{table*}[htp]
\caption{Polarization parameters of the IP and the MP for the mean pulse , the strong mode and the weak mode.}
 \centering
 \scriptsize
\begin{tabular}{cccccccccccc}
 \hline
State & \, &\,  &MP  &\, &\, &\, &\, &\, &IP &\,  &\, \\
\, &$\rm W_{50}$  &$\rm W_{10}$ &$\rm \langle L\rangle/I$ &$\rm\langle V \rangle/I$ &$\rm\langle \vert V \vert \rangle/I$ &\, &$\rm W_{50}$  &$\rm W_{10}$ &$\rm \langle L\rangle/I$ &$\rm\langle V \rangle/I$ &$\rm\langle \vert V \vert \rangle/I$  \\
\, &(Degree) &(Degree) &(Percent) &(Percent) &(Percent) &\, &(Degree) &(Degree) &(Percent) &(Percent) &(Percent) \\
 \hline
Mean &$9.3$ &$18.3$  & $69$  & $-1.3$  & $2.4$ &\, &$3.3$ &$10.9$ &$92$  & $-5.4$  &$5.4$ \\
Strong State &$9.3$ &$17.9$ &$70$  & $-1.5$  &$2.4$ &\, &$3.3$ &$10.9$ &$91$  & $-5.3$  &$5.3$ \\
Weak State &$8.8$ &$19,7$ &$62$  & $0.3$  & $2.1$ &\, &$3.0$ &$3.2$ &$99$  & $-11$  &$13$ \\ 
 \hline
 \end{tabular}
 \label{parameter}
\end{table*}

\subsection{The pulse profile evolution with frequency}
\label{sec:frequency}
Many pulsars exhibit significant evolution in their pulse shapes with observing frequency. The changes in pulse width and separation between different components are generally taken to be a consequence of emission from different heights \citep{1970_Craft,1978_Cordes_region}. Some other changes such as relative intensity differences between components are usually related to the variation in geometry or the emission mechanism\citep{2004_Lorimer_Handbook}.

\begin{figure}
\centering
\includegraphics[width=0.48\textwidth]{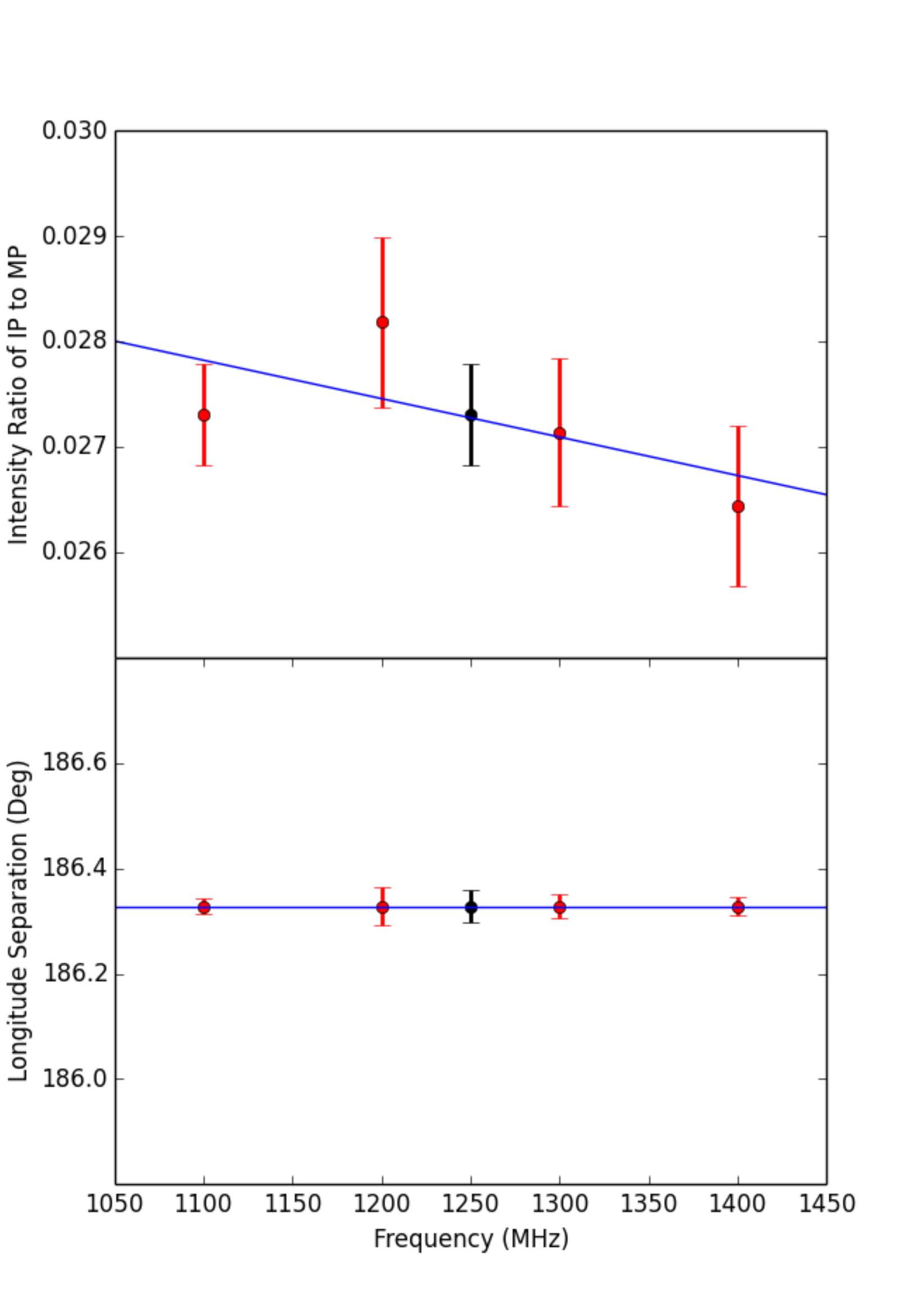}
\caption{The intensity ratio (top panel) and longitude separation (bottom panel) between the IP and MP at  different frequencies. The red points in each panel are the average values at the center frequencies of different frequency intervals, and the black points are those of all observing frequencies. The blue lines are best linear fits for the intensity ratio and the longitude separation as a function of frequency.}
\label{fig:RS}
\end{figure}

Fig. \ref{fig:RS} presents the intensity ratio (top panel) and longitude separation (bottom panel) between the IP and the MP at different frequencies. The best linear fits for the intensity ratio shows that the peak intensity ratio of the IP to the MP decreases as the frequency increases. However, the evolution trend is not clear because of the large measurement errors.

In the work of \citet{1986_Hankins}, a Gaussian curve was fitted to the weak IP, and the centroid was used as the IP fiducial point to measure its position from the peak of the MP. The separation between them was reported to be $187.4\pm0.2^\circ$, independent of frequency. However, the IP profile can be detected with high signal-to-noise ratio by the high-sensitivity observations of FAST, and the peaks of the IP and the MP were chosen to measure their separation in this paper. The errors are calculated based on the rms noise. The longitude separations between the IP and the MP at different frequencies are shown in the lower panel of Fig. \ref{fig:RS}.  It is $186.3\pm0.3^\circ$ in our work.  The separation is found to be independent of radio frequency, which is consistent with the results of  \citet{1986_Hankins}.

\section{Discussion}

\subsection{The periodic amplitude modulation}
\begin{figure}
\centering
\includegraphics[width=0.48\textwidth]{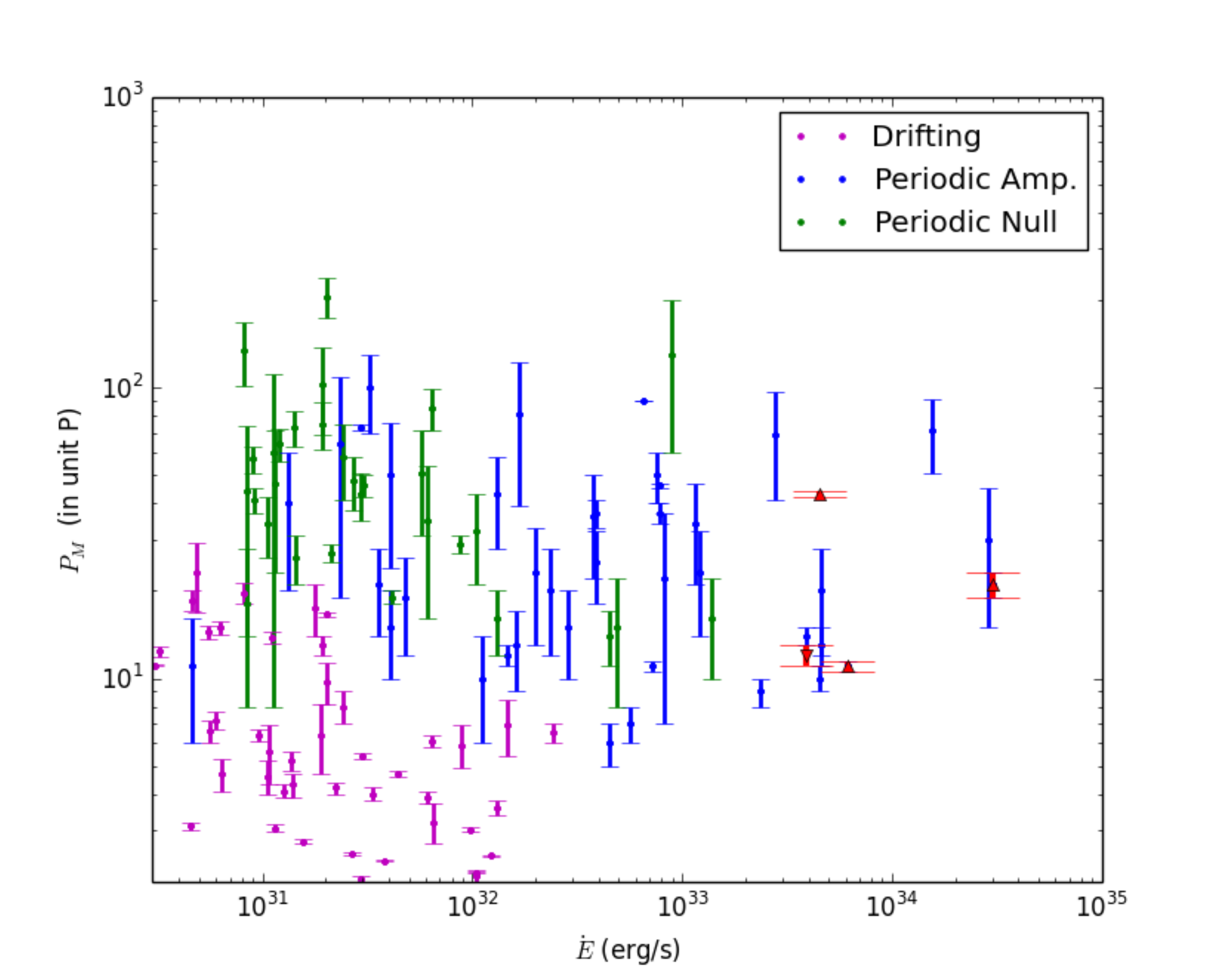}
\caption{The modulation periods ($P_M$) as a function of the spin-down energy loss ($\dot{E}$) \citep{2017_Basu_nulling,2019_Basu_dirfting,2020_Basu}. 
The magenta points are pulsars with subpulse drifting, the blue ones are for periodic amplitude modulation, and the green ones are for  periodic nulling. PSRs B1055-52,B1702-19 and B1822-09 are labelled as red triangles, and PSR B1929+10 is labelled as red inverted triangle.}
\label{fig:compare}
\end{figure}

The emission properties of PSR B1929+10 exhibit great complexity.  The most significant feature is the periodic modulation in the IP and parts of the MP. The weak preceding component of the MP shows the same modulation period as the IP (Fig. \ref{fig:FS2}). As shown in the left column of Fig. \ref{fig:FS1}, the fluctuation spectrum of the MP shows a broad low-frequency structure with two relatively significant features at $0.1 \, \rm cpp$ and $0.18\, \rm cpp$, mainly from its first two components. We confirm that it is periodic amplitude modulation rather than subpulse drifting because there is no systematic drifting pattern detected in the data. 
Recently, periodic amplitude modulation has been detected in $19$ radio pulsars \citep{2020_Basu,2020_Yan_1048-5832}.  \citet{2019_Yan_1822} reported that the IP and the precursor of PSR B1822-09 are modulated with a period of $\sim 43 P$ in its  quiet (Q) mode. Similar amplitude modulations were also found PSRs B1946+35, J1048-5832 and the burst(B) mode of PSR B0823+26 \citep{2017_Mitra_B1946+35,2020_Yan_1048-5832,2019_Basu_0823}.
\citet{2016_Basu_Amp_flu,2020_Basu} studied the known pulsars with periodic modulation (including systematic drifting and amplitude modulation), and found  clear differences between pulsars with subpulse drifting and those with periodic amplitude modulation/nulling, which are shown in Fig. \ref{fig:compare}. The subpulse drifting is only seen in pulsars with spin-down energy loss rate ($\dot{E}$) below $ 5 \times 10^{32} \, \rm erg /s$. However, periodic amplitude modulation can be seen in pulsars with relatively longer modulation period and larger spin-down energy loss rate, which implies that it has a distinct physical origin compared with subpulse drifting \citep{2020_Basu}. PSR B1929+10 has a relatively large spin-down energy loss rate of $\dot{E} \sim 3.93\times10^{33} \, \rm erg /s $, and modulation period $\sim 12 P$. It lies in the group with periodic amplitude modulation (the red inverted triangle in Fig. \ref{fig:compare}). 

Though there is no obvious boundary between the periodic amplitude modulation and periodic nulling, they are indeed different in their physical properties \citep{2019_Basu_dirfting}.  Similar to PSR B1822-09, there is no nulling pulse detected in the data of PSR B1929+10. We also made an on-pulse energy distribution for the IP, and found that the peak of on-pulse energy deviates from the zero point in the histogram, which confirmed that the periodic modulation of PSR B1929+10 is caused by emission mode switching.  

\subsection{The phase-locking features in other pulsars}
\label{sec:features}
The second remarkable feature in PSR B1929+10 is the phase-locked (delay) modulation between the IP and the MP. As analysed in subsection \ref{phaselock}, the $\sim 12  P$  modulation pattern in the IP is correlated (or anti-correlated) with that in the components of the MP. The energy fluctuations in the IP and the weak preceding component of the MP are anti-correlated. However, it is positively correlated with that of the first two components of the MP. What is more interesting is that the modulation of $\rm MP\_\uppercase\expandafter{\romannumeral2}$ is delayed about $1 P$. 
The phase-locked (delay) modulation is also detected in the observation on 2018 September $5$. Combined with the periodic amplitude fluctuation reported by \citet{2016_Basu_Amp_flu}, it is possible that the phase locking is a permanent feature for PSR B1929+10. It is puzzling that (1) there is a correlation transition from negative to positive on the leading edge of the MP, (2) there is a phase-locked (delay) between the IP and different parts of the MP.

Similar phase-locked (delay) modulation has been reported for PSRs B1702-19, B1055-52 and B1822-09 \citep{2007_Weltevrede_1702,2012_Weltevrede_1055,2019_Yan_1822}. For PSR B1702-19, its IP and the trailing part of the MP are modulated with a period of $\sim 10 \,P$, and are phase-locked by half-period delay \citep{2007_Weltevrede_1702}. Both the MP and the IP of PSR B1055-52 contain several emission components, and are modulated by a period of  $\sim 20 P$. Furthermore, the periodicity exhibits a phase-locked delay by $2.5\, P$ between the MP and the IP \citep{2012_Weltevrede_1055}. For PSRs B1702-19 and B1055-52, both of their modulation patterns are positive correlated. PSR B1822-09 was reported to switch between its B-mode and Q-mode according to the appearance and disappearance of the precursor component. Meanwhile, the IP changes in the opposite way to the precursor component. The average time between mode changes was a few minutes \citep{2019_Yan_1822}. However, there was no phase-locked (delay) modulation reported between the precursor and the IP.  A recent investigation revealed that the IP and the leading component of the MP were modulated with a period of $\sim 43 \,P$ in the Q-mode \citep{2019_Yan_1822}. This periodic modulation was found to be phase-locked.

The modulation parameters of PSRs B1055-52, B1702-19, B1822-09 and B1929+10 are shown in Table \ref{comparameter}. The comparison with other periodic modulation pulsars is shown in Fig. \ref{fig:compare}. It seems that pulsars with phase-locked (delay) modulation have relatively larger energy loss rates.

\begin{table}
\centering
\caption{The (modulation) parameters of PSRs B1055-52,B1702-19, B1822-09 and B1929+10.}
 \scriptsize
\begin{tabular}{ccccccc}
\hline
PSRB     &$P$           &$\dot{P}$                 &$P_{3}$   &$\dot{E}$                    &$P_{delay}$       &$\tau_{delay}$\\
\,          & \,         &($10^{-15}$)                    &\,                 &($10^{33}$)                 &\,                 &\, \\
\,          &$s$                 &$s/s$             & $P$             &$erg/s$       & $P$                      & $s$      \\
\hline
1055-52 &$0.197$ &$5.83$                      &$21\pm2 ^{a}$     &$30$                              &$2.5^{a}$               &$0.49$ \\
1702-19 &$0.299$   &$4.14$                       &$11\pm0.4^{b}$      &$6.1$                           &$0.5^{b}$                &$0.15$\\
1822-09 &$0.769$ &$52.5$                      &$43\pm1^{c}$      &$4.5$                         &$-$                   &$-$ \\
1929+10 &$0.227$ &$1.16$                       &$12\pm1$     &$3.9$                          &$1$                    &$0.23$\\
\hline
\end{tabular}
\begin{tablenotes}
        \footnotesize
        \item a,\citep{2007_Weltevrede_1702}.
        \item b,\citep{2012_Weltevrede_1055}.
        \item c,\citep{2019_Yan_1822}.
\end{tablenotes}
\label{comparameter}
\end{table}

For PSR J0826+2637, its MP and postcursor component  also show periodic amplitude modulations in the B-mode. However, the IP is too weak to detect any modulation features \citep{2019_Basu_0823}. In addition to the periodic modulation in the intensity, the pulse width of single pulses also shows periodic changes at both the leading and trailing sides in its B-mode \citep{2019_Basu_0823}. Similar to PSR J0826+2637, the MP's pulse width of PSR B1929+10 also changes periodically at its leading side due to the intensity fluctuation of $\rm MP\_\uppercase\expandafter{\romannumeral2}$. 

\subsection{The emission geometry for PSR B1929+10}
Much efforts have gone into studying the emission mechanism and magnetospheric structure of the pulsar. It seems that the relatively small inclination angle has categorised PSR B1929+10 as a single-pole interpulsar. Furthermore, the weak ``bridge'' emission between the MP and the IP provides important evidence for the single-pole emission hypothesis \citep{1977_Manchester,1985_Gil_interpulse}. However, the single wide cone model of \citet{1977_Manchester} is first ruled out in the case of PSR B1929+10, because of the frequency independence of the separation between the IP and the MP. In the single-pole scenario, \citet{2005_Dyks} proposed a competing model to explain the anti-correlation between the energy fluctuation in the IP and the precursor component of PSR B1822-09. It assumes that there is an emission region close to the MP emission region, which changes its emission direction outward/inward towards the neutron star, and would be respectively observed as the precursor component and the IP as the pulsar rotates. This model suggests that the precursor component and the IP should have the same origin but outside the region of the MP. As discussed in subsection \ref{sec:features}, while there are many observational similarities between PSRs B1822-09 and B1929+10, the anti-correlation between the weak preceding component of the MP and the IP could be understood in the same way. However, a separation of $180^\circ$ between the the anti-correlated components is expected in this model, which is obviously not the case for PSR B1929+10, a separation of $211^\circ$ is measured between the IP and the weak preceding component. \citet{2007_Weltevrede_1702} discussed a possibility caused by retardation and aberration in a bidirectional model, which says that emission from higher would arrive earlier to the observer. According to this model, the emission height of the IP and  $\rm MP\_\uppercase\expandafter{\romannumeral1}$ is expected to be $700\, \rm km$ if the minimum of the linear polarization  between $\rm MP\_\uppercase\expandafter{\romannumeral1}$ and $\rm MP\_\uppercase\expandafter{\romannumeral2}$ is setted to be the fiducial points (longitude of the magnetic pole on the star). The derived height of the MP is expected to be $850\,\rm km$ and its emission beam defined by the last closed fieldlines is $24^\circ$. Accordingly, the component width is expected to be $42^\circ$ which is roughly consistent with the observed width of the MP.

The polarization profile of PSR B1929+10 has been studied in detail, showing that both the center of the MP and the IP have relatively higher circular polarization, which indicates that both the MP and the IP
should be dominated by core emission \citep{1983_Rankin_core,1997_Rankin_1929,2015_Basu_PPC}. By identifying the core components and calculating their widths from the MP and the IP regions, PSR B1929+10 has been taken as an orthogonal rotator since then. In the two-pole scenario, both poles of the pulsar are synchronously modulated with the same period of $\sim 12 P$. Just like PSR B1055-22 \citep{2012_Weltevrede_1055}, it seems that the MP of PSR B1929+10 imitates the IP in intensity and pattern. There must be information transmission from one pole to the other. \citet{2012_Weltevrede_1055} discussed possible physical mechanisms of neutron star nonradial oscillation, which provided a way to explain the interpole phase-locking via the neutron star surface. The positive correlation and the $\sim 1 P$ modulation delay between the IP and the $\rm MP\_\uppercase\expandafter{\romannumeral2}$ can be naturally explained in this model. However, we still need to explain the  anti-correlation between the modulations in the IP and the weak preceding components of the MP.

Compared with postcursor and precursor components of other pulsars, the PC of PSR B1929+10 is the widest of all, being  about $\sim 96^{\circ}$. It is fully linearly polarized and there is no modulation feature found in this component. Moreover, its position also stay constant with frequency (see the bottom panel of Fig. \ref{fig:bridge}). \citet{2015_Basu_PPC} summarized the pre/post-cursor emission of different pulsars, and demonstrated the pre/post-cursor should have an origin outside the region of the MP and have a different emission mechanism.

To sum up, the interaction between the radio emission from the MP and the IP is a conundrum for present pulsar theories, and has important implications for understanding the emission mechanism and structure of the pulsar magnetosphere. According to the emission properties, all $44$ known interpulse emitting pulsars are classified into two groups: double-pole model and single-pole model \citep{2011_Maciesiak_IP}.
Evidence for alignment of the magnetic axis with the rotational axis was found by comparing the characteristic ages of pulsars in these two groups \citep{2011_Maciesiak_IP}. 

\section{Conclusions}
We have carried out a detailed single pulse analysis on PSR B1929+10 based on FAST observations. Our high sensitivity observation reveals that the weak IP shows clear pulse-to-pulse modulation. With the fluctuation spectrum analysis, we found that both the MP and IP are modulated by a period of $\sim 12 P$, and confirmed that it is a periodic amplitude modulation because there are no obvious subpulse drifting features detected in our data. Due to the high time resolution possible with the FAST sensitivity, the components which contribute to the $\sim 12 P$ modulation could be well identified. Correlation studies revealed that the fluctuation in the IP is anti-correlated with that in the weak preceding component of the MP ($\rm MP\_\uppercase\expandafter{\romannumeral1}$), but correlated with that in the first two components of the MP ($\rm MP\_\uppercase\expandafter{\romannumeral2}$). We also found that the modulation in the first two components of the MP is delayed compared to that in the IP by about $1P$. We classified the pulses as to being in strong and weak modes according to the periodic energy variation in the IP. The pulse profile shape of the MP also changes because parts of its components have the same modulation period as the IP. It is puzzling that the IP and the MP are phase locked. These facts imply that the magnetosphere should be modulated periodically and globally, and there should be transfer of information between different emission regions or poles. 

Weak ``bridge'' emission between the IP and the MP is clearly detected at $1.25\, \rm GHz$, which has been taken as important evidence favoring the single-pole model.  
However, the separation between the IP and the MP is confirmed to be independent of observing frequency, which rules out the single wide cone model in the single-pole scenario. Observations of PSR B1929+10 with FAST represent a conundrum for pulsar theories and cannot be satisfactorily explained by the current pulsar models. Much work still needs to be done to fully understand
pulsar radio emission and the structure of neutron star magnetosphere.

\section{ACKNOWLEDGEMENTS}
We would like to thank the referee for the comments and suggestions which helped to improve the paper.
This work is supported by the West Light Foundation of the Chinese Academy of Sciences (No. 2018-XBQNXZ-B-023), the National Key R$\&$D Program of China under grant number $2018\rm YFA 0404703$, the Open Project Program of the Key Laboratory of FAST, NAOC, Chinese Academy of Sciences, the China Postdoctoral Science Foundation grant ($2019\rm M 650847$) and the Cultivation Project for FAST Scientific Payoff and Research Achievement of CAMS-CAS. WMY is supported by the 2021 open program of the Key Laboratory of Xinjiang Uygur Autonomous Region, the NSFC (Nos. $\rm U1831102$, $\rm U1731238$, $\rm U1838109$, $\rm U11873080$) and the CAS ``Light of West China'' Program (No. 2017-XBQNXZ-B-022). RY is supported by the West Light Foundation of the Chinese Academy of Sciences, project (No. 2016-QNXZ-B-24).YZY is supported by the Guizhou Education Department, grant (No. Qian Education Contract KY[2019]214). DL is supported by CAS Strategic Priority Research Program (No. XDB23000000). YLY is supported by the National Key R$\&$D Program of China (2017YFA0402600) and the CAS ``Light of West China'' Program.

The FAST FELLOW-SHIP is supported by Special Funding for Advanced Users, financed and administered by the Center for Astronomical Mega-Science, Chinese Academy of Sciences (CAMS). This work made use of the data from FAST (Five-hundred-meter Aperture Spherical radio Telescope).  FAST is a Chinese national mega-science facility, operated by the National Astronomical Observatories, Chinese Academy of Sciences.
 

\bibliography{tex}

\end{document}